\documentclass[prx,aps,showpacs,twocolumn,preprintnumbers,
amsmath,amssymb,superscriptaddress,longbibliography]{revtex4-1}
\usepackage[english]{babel}
\usepackage{amsmath}
\usepackage{amssymb}
\usepackage{amsfonts}
\usepackage{physics}
\usepackage{graphicx}
\usepackage[colorlinks=True,linkcolor=red,citecolor=Red,urlcolor=Red]{hyperref}
\usepackage{bookmark}
\usepackage[dvipsnames]{xcolor} 
\usepackage{braket}
\usepackage{nicefrac}
\usepackage{bm}
\usepackage{bbm}
\usepackage{braket}
\usepackage{slashed}
\usepackage{mathdots}
\usepackage[normalem]{ulem}
\usepackage{array}
\usepackage{makecell}

\usepackage{cmap}
\usepackage{setspace}
\usepackage{amstext}			% \text{}
\usepackage{amsfonts}
\usepackage{bbold}
\usepackage{graphicx}
\usepackage{subfigure}
\usepackage{wrapfig}
\usepackage{esint}
\usepackage{hyperref}
\usepackage{multirow}
\usepackage{printlen}
\usepackage{placeins}
\usepackage{comment}
\usepackage{bm}
\usepackage{tikz-cd}
\usepackage{tkz-berge}
\usepackage[boxsize=1em]{ytableau}
\usepackage[english]{babel}
\usepackage{fancyhdr}
\usepackage[left=1in, right = 1in, top = 1in, bottom = 1in]{geometry}

  \DeclareUnicodeCharacter{2212}{-}

\DeclareMathOperator{\im}{Im}

\DeclareMathOperator{\ord}{ord}
\DeclareMathOperator{\trop}{trop}

\newcommand{\ttr}[1]{{\rm tr}{\left[#1\right]}}

\newcommand{\BH}{\mathcal{H}}

\newcommand{\BP}{\mathcal{P}}

\newcommand{\BL}{\mathcal{L}}

\newcommand{\BF}{\mathcal{F}}

\newcommand{\id}{\mathbb{1}}

\allowdisplaybreaks

\begin{document}

 \title{
Characterizing all non-Hermitian degeneracies 
using 
algebraic 
approaches:
Defectiveness and asymptotic behavior 
 }
  \author{Sharareh Sayyad}
 \affiliation{Department of Mathematics and Statistics, Washington State University, \\Pullman, Washington 99164-3113, USA}
 \affiliation{Institute for Numerical and Applied Mathematics, Georg-August University of G\"ottingen, Bunsenstra\ss e 3-5, G\"ottingen 37073, Germany}
 \email{sharareh.sayyad@wsu.edu}
 \author{Grigory A. Starkov}
  \affiliation{Institute for Theoretical Physics and Astrophysics,
University of W\"urzburg, D-97074 W\"urzburg, Germany}
\affiliation{W\"urzburg-Dresden Cluster of Excellence ctd.qmat, Germany}
\email{grigorii.starkov@uni-wuerzburg.de}

\begin{abstract}
Degeneracies in Non-Hermitian (NH) systems range from Hermitian-like $n$-bolical points to exceptional points~(EPs) with different Jordan structures. When several Jordan blocks share the same eigenvalue, multi-block degeneracies form, whose perturbative behavior has so far lacked systematic analysis. In this work, we present an algebraic framework based on tropical polynomials and their equivalence to Newton polygons to characterize the asymptotic splitting of all types of NH degeneracies under both generic and non-generic perturbations. By tropicalizing the characteristic polynomial, derived from traces of matrix powers without any matrix diagonalization, we reproduce the Lidskii–Vishik–Ljusternik asymptotics for generic perturbations. Additionally, we uncover anomalous exponents, such as a 2/3 splitting of a two-block EP, along non-generic perturbation directions. We further tabulate the complete splitting behavior for all Jordan structures of matrices up to size four.
Finally, using a range of physically motivated systems, including EP-based sensors, the non-Hermitian Lieb model, and multi-block EPs of Liouvillian superoperators, we demonstrate that our algebraic approach facilitates the analysis of NH degeneracies across various experimentally relevant settings.

\end{abstract}

 \maketitle

\section{Introduction}

In recent years, we have witnessed a surge of interest in non-Hermitian~(NH) physics~\cite{Ashida2020non, guriasone2020exceptional, Wang_2021, Ding2022non}, which describes systems exchanging particles or energy with their environment, as well as effective descriptions of subsystems embedded in larger closed systems. One of the distinct properties of NH systems, without counterparts in Hermitian physics, is the possibility of hosting exceptional points~(EPs), spectral degeneracies at which eigenvalues and eigenvectors coalesce. Exploring the behavior of NH systems near EPs has yielded rich results from both theoretical and practical perspectives: the emergence of eigenvalue braiding and its exotic topology~\cite{Wang2021, Patil2022, Liang2024, Guria2024}, the appearance of EPs on phase boundaries of coupled and correlated systems~\cite{Sayyad2021, Sayyad2022b, Gal2023, Lado2023, Zhang2025b}, the enhancement of sensor sensitivity~\cite{chen2017exceptional, Khajavikhan-17, lai2019observation, kononchuk2022exceptional, Parto2025enhanced, Behrouzi-25, Chowdhury-25, Zheng-25, Wiersig-26}, and the acceleration of entanglement generation near EPs~\cite{Li2023entanglement} exemplify these developments.

The coalescence of eigenvalues in NH models is classified by the Jordan canonical form in two independent respects. (i) An NH degeneracy with algebraic multiplicity $n>1$ is represented by a set of Jordan blocks whose sizes sum to $n$~\cite{KostrikinManin}. If all blocks have size one, the degeneracy is \emph{nondefective} and Hermitian-like; such degeneracies are known as $n$-bolical points. However, when at least one block has size larger than one, the matrix is nondiagonalizable, and the degeneracy is \emph{defective}; defective degeneracies are the EPs. (ii) If a single Jordan block of size $n$ represents the eigenvalue, the degeneracy is \emph{non-derogatory}. \emph{Derogatory} degeneracies emerge when several Jordan blocks represent the same eigenvalue~\footnote{Thus, $n$-bolical points are an extreme case of this type of derogatory degeneracy, with $n$ unit blocks.}. Derogatory degeneracies have also been referred to as fragmented~\cite{bid_PhysRevResearch, bid_2025} or multiblock~\cite{shiralieva2025} EPs. Their response to perturbations of a physical parameter has remained largely unexplored; addressing it is the focus of the present work. Throughout, we assume that there is only a single degenerate eigenvalue whose algebraic multiplicity coincides with the matrix dimension; several degeneracies at different eigenvalues reduce to this case by projection onto the relevant generalized eigenspace.

Exploration of EPs in the literature primarily focuses on three main research themes. (i) Classification: identifying all possible degeneracies an NH matrix can host, whether the matrix is an NH Hamiltonian~\cite{Yoshida2018, Michishita2020, Crippa2021, Erdamar2025exploring} or the vectorized form of a Liouvillian super-operator~\cite{Minganti2019, Chen2021b, Sayyad2021, Hamanaka2023, Erdamar2024}. 
(ii) Perturbative response: characterizing the asymptotic behavior of EPs upon varying a tuning parameter~\cite{patil2022measuring, Wingenbach2024, zhang2025dynamically, Wiersig2025} and, in particular, addressing how eigenvalues split under perturbation.  
(iii) Methodology: developing intuitive and computationally stable formulations to theoretically tackle (i) and (ii), for instance, methods based on discriminants and resultants~\cite{Delplace2021, Sayyad2022, Yoshida2022discriminant, Sayyad2022b, Sayyad2023, stalhammar2025}.
These methods, however, were developed to characterize non-derogatory EPs and do not extend to derogatory degeneracies. 
In the present work, we address all three themes by classifying all types of NH degeneracies and determining their splitting asymptotics under both generic and non-generic perturbations, using a tropical-geometric approach that is mathematically rigorous and yet simple. The complementary question, namely which degeneracy types can be converted into one another by infinitesimal perturbations without lifting the degeneracy, is addressed in our recent paper~\cite{Starkov2026}.

Investigating the behavior of an NH matrix under a small perturbation parametrized by $\epsilon$ was the focus of many studies on non-derogatory EPs~\cite{Budich2020, Mandal2021, Sayyad2022, Sayyad2023, Banerjee2023, Bid2025}. In many such studies, the appearance of $\epsilon^{1/q}$ dispersion for eigenvalues, with $q$ a positive integer, was associated with EPs of order $q$. There have also been examples of EPs of order $q$ whose eigenvalues disperse as $\epsilon^{1/(q-l)}$, where $l$ is a positive integer less than $q$~\cite{Mandal2021, Sayyad2022, Wang2025ep}. Most of these results have been obtained through case studies on non-derogatory EPs; recently, response theory has extended the leading-order analysis to derogatory EPs under generic perturbations~\cite{bid_PhysRevResearch, bid_2025}. A systematic exploration of all possible dispersions under both generic and non-generic perturbations, for all types of NH degeneracies, has nevertheless remained absent from the literature. Indeed, for generic perturbations, such splittings are governed by the classical Lidskii-Vishik-Ljusternik theory~\cite{Lidskii-1966, Moro-2003}; non-generic perturbations and derogatory structures fall outside its standard scope. Here, we introduce a systematic formulation that reproduces the generic asymptotics and extends them to non-generic perturbations and derogatory structures alike, providing complete results for perturbed square matrices of sizes 2, 3, and 4.

A natural language for such an exploration comes from algebraic geometry. Algebraic approaches, in particular amoeba theory, have recently advanced the field of the NH skin effect~\cite{Wang2024amoebaprx, Wang2024amoeba, hu2025topological, Yang2025amoeba, Kaneshiro2025}, a phenomenon in which a large number of eigenvectors localize at the boundaries of the system. Mathematically, in certain limits, amoebas~\footnote{An amoeba is the image of a polynomial's zero set under the logarithmic map~\cite{mikhalkin2001amoebas}.} reduce to their spines, also known as tropical diagrams or tropical curves~\cite{mikhalkin2001amoebas}. These objects need not be reached through amoebas, however: they are native to the field of tropical geometry, where redefined addition and multiplication operations map a polynomial to a tropical polynomial, a piecewise-linear function whose graph is the tropical diagram. Ref.~\cite{Banerjee2023} demonstrated, on a handful of examples, that tropical polynomials can be used to explore NH degeneracies. A closely related description is the Newton polygon, which arises naturally in constructing the series expansion of the eigenvalues in the perturbation parameter and encodes the same information as the tropical polynomial. It is precisely this correspondence that we exploit: we present a self-contained account of tropical polynomials, clarify their equivalence to Newton polygons, and use them to characterize all NH degeneracies systematically. Concretely, we obtain (a) complete splitting tables for all Jordan structures up to size four, under generic and non-generic perturbations; (b) anomalous exponents such as a $2/3$ splitting of a two-block EP, invisible to generic-perturbation theory; (c) validation on various physical systems.

The present work is arranged as follows. In Sec.~\ref{sec:trop-poly}, we discuss the eigenvalue perturbation problem and introduce the tools necessary for its analysis: Newton polygons and tropical polynomials. 
In Sections~\ref{sec:2x2},~\ref{sec:3x3} and~\ref{sec:4x4}, we analyze the eigenvalue splitting for both generic and non-generic perturbations for systems described by square matrices of size $2$, $3$ and $4$, respectively. In Sec.~\ref{sec:examples}, we apply the results from previous sections to physically motivated examples. Finally, we conclude this paper in Sec.~\ref{sec:conclusion}.

\section{Eigenvalue perturbation problem: Newton polygons and Tropical polynomials}
\label{sec:trop-poly}

\subsection{Leading order behavior and Newton polygon}

Understanding the behavior of the eigenvalues of a $n\times n$ matrix $A_{0}$ under a small perturbation~($\epsilon B$) such that~\cite{Moro-2003}
\begin{equation}
A(\epsilon) = A_0 + \epsilon B,
\end{equation}
can be equivalently expressed as characterizing the roots of the corresponding characteristic polynomial given by
\begin{equation}
\BF_\lambda(\epsilon) = \lambda^n 
+\dotsc + a_i(\epsilon)\lambda^{n-i} +\dotsc + a_n(\epsilon).
 \end{equation}
Here, $B$ denotes the perturbation matrix, and $\epsilon\in\mathbb{C}$ is the perturbation parameter.

Without loss of generality, we can assume that $A_0$ corresponds to a single eigenvalue $\lambda$ with algebraic multiplicity $n$ and these eigenvalues can vanish upon subtracting $\tr[A_{0}]/n$. The coefficients $a_i(\epsilon)$ are then some polynomials in $\epsilon$ with $a_1(0)=0$.

Let $\alpha_i$ be the smallest power of a term present in $a_i(\epsilon)$. If we are interested only in the leading-power behavior of the eigenvalues, we can substitute the ansatz 
\begin{equation}
\lambda(\epsilon) = \lambda_0 \varepsilon^\beta + {\cal O}(\varepsilon^\beta),
\end{equation}
into the equation $\BF_\lambda=0$ and then only keep the lowest-order terms. The leading power of the term $a_i(\varepsilon)\lambda^{n-i}$ reads
\begin{align}
    \mathrm{ord}_i=\alpha_i + \beta (n-i),
    \label{eq:lowest_ord}
\end{align}
with $\alpha_0=0$.
To have a non-trivial solution, we should have at least two lowest-order terms; say $\mathrm{ord}_i$ and $\mathrm{ord}_j$ such that all the other lowest-order terms have higher orders.

Rephrasing these statements reads as follows. There are at least two indices $i$ and $j>i$, for which $\mathrm{ord}_i=\mathrm{ord}_j$ and for all other indices $k$, we have $\mathrm{ord}_k\geqslant \mathrm{ord}_i$.
Rearranging Eq.~\eqref{eq:lowest_ord} for indices $\{i,j,k\}$, we get
\begin{align}
    \alpha_j-\alpha_i & =\beta (j-i),\label{eq:low-ord1}\\
    \alpha_k - \alpha_i& \geqslant \beta (k-i),\label{eq:low-ord2}\\
    \alpha_k - \alpha_j& \geqslant \beta (k-j)\label{eq:low-ord3}.
\end{align}
These conditions can be visualized geometrically with the help of the so-called {\it Newton polygon}: First, we plot the set of two-dimensional points $(i, \alpha_i)$. Then, we draw the lower boundary of the convex hull of this set of points, resulting in a polygon.
Finally, conditions~\eqref{eq:low-ord1}--\eqref{eq:low-ord3} imply that the leading-order power exponent $\beta$ is one of the slopes of this Newton polygon. In addition to that, we can also determine the number of eigenvalues corresponding to the leading power exponent $\beta$. If the Newton polygon has slope $\beta$ between horizontal coordinates $j$ and $i$, then the number of the corresponding eigenvalues is $|j-i|$.

In principle, the shape of the Newton polygon depends on the concrete structure of both the initial matrix $A_0$ and the perturbation matrix $B$. Nevertheless, for a general $B$ which completely lifts off the degeneracy, Newton polygon is determined solely by the Jordan block structure of $A_0$~\cite{Moro-2003}.

The asymptotic behavior of the eigenvalues is then given by the Lidskii or Lidskii-Vishik-Ljusternik theorem~\cite{Lidskii-1966, Moro-2003}, which states as follows. Assume that the degenerate eigenvalue of our system has $r_j$ Jordan blocks of size $n_j$ for $j=1,2,\dotsc,q$, where $n_1> n_2>n_3\dotsc>n_q$. Then, for a generic perturbation (with generic leading powers of the characteristic polynomial coefficients), there are $r_j$ series of $n_j$ eigenvalues admitting the first-order expansion
\begin{equation}
\lambda^{kl}_j(\varepsilon) = (\xi^k_j)^{1/n_j} \varepsilon^{1/n_j}e^{i\frac{2\pi (l-1)}{n_j}} + {\cal O}(1/n_j),
\label{lidskii-result}
\end{equation}
where $k=1,2,\dotsc,r_j$ labels the series and $l = 1,2,\dotsc,n_j$ sets the index of the eigenvalue inside the series. 
The parameters $\xi^k_j$ are the roots of the certain characteristic equations for the matrices constructed out of $B$ and Jordan basis vectors of matrix $A$~\footnote{ We will not further go into details. Interested readers are encouraged to consult Refs.~\cite{Lidskii-1966, Moro-2003}.}.

As we see in Eq.~\eqref{lidskii-result}, the eigenvalues with different $l$ correspond to the different complex branches of $\varepsilon^{1/n_j}$.
We can employ Eq.~\eqref{lidskii-result} to analyze the braiding of eigenvalues along the infinitesimal loop $\varepsilon(\varphi) = \varepsilon_0 e^{i\varphi}$. The $r_j$ Jordan sub-blocks of size $n_j$ produce $r_j$ series of $n_j$ eigenvalues. For each of the series, braiding corresponds to the cyclic permutation of eigenvalues.

Finally, we should point out that these results are merely applicable to the one-parameter perturbations. In the cases we consider in the paper, the space of the perturbations is typically multi-dimensional. However, we can always reduce the problem to one-dimensional if we restrict the perturbation parameters to complex linear one-dimensional subspace. Such a procedure is essentially equivalent to fixing the perturbation matrix $B$.

\subsection{Tropical Polynomials}\label{sec:troppol}

As we have noted, the information about the leading-power behavior of the eigenvalues under perturbation is reflected in the leading powers of the coefficients of the characteristic polynomial. There are other equivalent ways to represent this information besides Newton polygons.
One such approach employs tropical geometry~\cite{Akian2006, Banerjee2023}.

Before we explain this approach in more detail, we should introduce the notion of the {\it min-plus semiring} $\mathbb{R}_\mathrm{min}$. The $\mathbb{R}_\mathrm{min}$ is the set $\mathbb{R}\cup\{\infty\}$ equipped with the operations of tropical addition $\oplus$ and tropical multiplication $\odot$~\footnote{We note that there is no subtraction in the tropical arithmetic. This is the reason for dealing with {\it semiring} $\mathbb{R}_\mathrm{min}$.} such that for $\{a,b\}  \in \mathbb{R}_\mathrm{min}$, we have
\begin{align}
a \oplus b & = \min\{a,b\},\\
a \odot b & = a+b,
\end{align}
The operations in $\mathbb{R}_\mathrm{min}$ mimic the behavior of the leading (smallest order) power exponents of polynomials (or function series expansions) when we add or multiply these polynomials (series expansions). In other words, the leading power exponent of $f(\epsilon)$, denoted by $\ord(f)$, should satisfy
\begin{align}
\ord(f+g) & \geqslant \min\{\ord(f),\ord(g)\},\\
\ord(f\times g) & = \ord{f} + \ord{g}.
\end{align}

If we have a polynomial $\BF_{\lambda}(\epsilon)$ in $\lambda$ that depends on the parameter $\epsilon$
\begin{equation}
\BF_{\lambda}(\epsilon) = \sum_{i=0}^{n} a_i(\epsilon)\lambda^{n-i},
\end{equation}
we can match it with the corresponding tropical polynomial in a new variable $\omega_{0} \in\mathbb{R}_\mathrm{min}$ such that
\begin{equation}
    \BP(\epsilon, \omega_{0}) = \trop(\BF_{\lambda})(\omega_{0}) = \bigoplus_{i=0}^n \ord{(a_i)}\omega_{0}^{n-i}.
\end{equation}
Here, all $\ord(a_i)$ are given in $\mathbb{R}_\mathrm{min}$, and the associated operations, such as summations and multiplications, are also defined in $\mathbb{R}_\mathrm{min}$. Translating it into a more readable language,
\begin{align}
    \ord{(a_i)}\omega_0^{n-i} &= \ord{(a_i)}\odot\underbrace{\omega_0\odot\omega_0\odot\dotsc\odot\omega_0}_{n-i} 
    \,, \nonumber \\
    &= \ord{a_i} + (n-i)\omega_0,
\end{align}
and
\begin{multline}
    \BP(\epsilon, \omega_{0}) = \min\{\ord{(a_0)}+n\omega_0, \\\ord{(a_1)}+(n-1)\omega_0, \dotsc,\ord{(a_n)}\}.
\end{multline}

Analogously to algebraic polynomials, tropical polynomials can be expanded in terms of products of monomials~\footnote{A monomial is any product of variables in tropical semiring $\mathbb{R}_\mathrm{min}$. Repetitions are allowed in monomials.} as~\cite{Akian2006} 
\begin{align}
\BP(\epsilon,\omega_{0}) = \ord(a_0)&\odot (\omega_{0}\oplus \omega_1)\odot (\omega_{0}\oplus \omega_2)\nonumber \\
&\odot \dotsc \odot (\omega_{0}\oplus \omega_n),\label{trop-expansion}
\end{align}
where $\omega_i$s are dubbed {\it tropical roots}, which correspond to the points of non-differentiability of $\BP(\epsilon, \omega_{0})$.

One may wonder how this abstract mathematical approach may unveil the behavior of the polynomial roots under perturbation. As it turns out, the tropical roots $\omega_i$ are precisely the leading power exponents of the roots of the original polynomial; see the proof by B. Sturmfels in Ref.~\cite{Sturmfels2007}.

To sketch the idea of the proof, we first check that it works for a monomial. The tropicalization of the polynomial
\begin{equation}
\BF_{\lambda}(\epsilon) = a_0(\epsilon)\lambda + a_1(\epsilon),
\end{equation}
results in
\begin{align}
\BP(\epsilon,\omega_{0}) 
&= 
\ord(a_0) \odot\omega_{0} \oplus \ord(a_1)   ,\nonumber \\
&= \min\{\ord(a_0)+\omega_{0}, \ord(a_1)\} 
,\nonumber \\
&=
\ord(a_0) + \min\{\omega,\ord(a_1)-\ord(a_0)\} 
,\nonumber \\
&=
\ord(a_0)\odot(\omega_{0} \oplus \ord(a_1/a_0)).
\end{align}

Secondly, one can show that $\trop(f\times g)=\trop(f)\odot\trop(g)$ (as functions of the new variable $\omega_{0}$). If we apply this property to the tropicalization of the monomial product expansion of the polynomial
\begin{equation}
\BF_{\lambda}(\epsilon) = a_0(\epsilon)(\lambda + \lambda_1(\epsilon))(\lambda + \lambda_2(\epsilon))\dotsc(\lambda+\lambda_n(\epsilon)),
\end{equation}
then we directly arrive at the proof. Here, $\lambda_i(\epsilon)$ are the roots of the polynomial with  inverted signs.

We can also extract the multiplicities of the tropical roots from the tropical polynomial in Eq.~\eqref{trop-expansion}. If the multiplicity of the root $\omega_i$ is $n_i$, then the change of the slope at the root is determined by $(\omega_{0} \oplus \omega_i)^{n_i} = n_i\times\min\{\omega_{0}, \omega_i\}$, i.e., it is equal precisely to $-n_i$.

\subsection{Characteristic polynomial and parameterization}

After introducing the characteristic polynomials and tropical geometric approaches, we proceed presenting the tropicalization of characteristic polynomials.

The characteristic polynomial for a matrix $\cal H$ of rank $n$ in general casts
\begin{align}
    {\cal F}_{\lambda}
    =
    \lambda^{n} - \sigma_{1} \lambda^{n-1}
    + \ldots
    +(-1)^{n} \sigma_{n},
    \label{eq:Fgen} 
\end{align}
where $\sigma_{k}$s are the trace of the $k$th exterior power of ${\cal H}$~\footnote{
The trace of the exterior power $\sigma_{k} = \tr(\wedge^k {\cal H})$
can be also expressed as $\sigma_{k}= (-1)^{-k} p_{k}$ with~\cite{Sayyad2022}
$ 
    p_{k} = 
    -\frac{s_{k} + p_{1} s_{k-1} +\ldots +p_{k-1} s_{1}}{k},$
     and $s_{k}=\tr[{\cal H}^{k}]$.
}, given by
\begin{align}
    \sigma_{1} = \tr[{\cal H}], 
    \, 
    \sigma_{n} = \det[{\cal H}], 
\end{align}
and 
\begin{align}
 \sigma_{k} &= \frac{1}{k!} \det \begin{bmatrix}
\tr [{\cal H}] & k-1 & \cdots&0  \\
\tr [{\cal H}^2] & \tr [{\cal H}] &  \cdots &0 \\
\vdots & \vdots & \ddots & \vdots \\
\tr[{\cal H}^{k-1}] & \tr[{\cal H}^{k-2}] & \cdots & 1 \\
\tr[{\cal H}^k] & \tr[{\cal H}^{k-1}] & \cdots & \tr [{\cal H}]
\end{bmatrix},
\label{eq:sigmak}
\end{align}
with $k \in \{ 2 ,\ldots ,n-1\}$. Without loss of generality, we keep ${\cal H}$ to be traceless and set $\sigma_{1}=0$. 

Before drawing a generic picture, we first explore a few special cases. 

When $\sigma_{n} \neq 0 $ and $\sigma_{i}=0$ with $i\in \{1, \ldots n-1\}$, the roots of the characteristic polynomial ${\cal F}_{\lambda}= \lambda^{n} +(-1)^{n} \sigma_{n} $ are $\lambda = -e^{2\pi i j/n}[(-1)^{n+1} \sigma_{n}]^{1/n} $ for $j=1, \ldots n$. To characterize the degeneracies of this case, we now consider three different situations. i) Considering $\sigma_{n} = \prod\limits_{i=1}^{n} \varepsilon_{i}$ with $\varepsilon_{i}$ being the $i$th eigenvalue of ${\cal H}$, we can introduce a linear 1D asymptotic expansion around one of eigenvalues, say $m$, as $\tilde{\sigma}_{n} (t)= (\varepsilon_{m} -\tilde{\varepsilon}_{m} t )\prod\limits_{i \neq  m } {\varepsilon_{i}}$. This results in ${\cal F}_{\lambda}( t)= \lambda^{n} +\tilde{\sigma}_{n} (t)  $. The associated tropical polynomial then casts ${\cal P}(t, \omega_{0})= \min\{ n \omega_{0}, 1\}$ with the tropical roots $\omega_{0}=1/n$.
Hence, these degeneracies are EP$n$s with $1/n$ asymptotic dispersion.
ii) Considering all eigenvalues to be identical~($\varepsilon_{i}=\varepsilon_{0}$) and their asymptotic expansions with parameter $t$ is also the same, we get $\tilde{\sigma}_{n} (t)= \prod\limits_{i=1}^{n} (\varepsilon_{i} -\tilde{\varepsilon}_{i} t ) = \sum\limits_{k=0}^{n} 
\begin{pmatrix}
    n \\
    k
\end{pmatrix}
(-1)^{k} \varepsilon_{0}^{n-k} \tilde{\varepsilon}_{0}^{k} t^{k}
$. As $\cal H$ is traceless, identical eigenvalues should be $\varepsilon_{0}=0$, i.e., diagonalized $\cal H$ is a null matrix. The associated tropical polynomial  then casts ${\cal P}(t, \omega_{0})= \min\{ n \omega_{0}, n\}$ with its root being $\omega_{0}=1$.
These degeneracies are non-defective degeneracies with linear asymptotic dispersion.
iii) However, when $\tilde{\varepsilon}_{0} = 0$, we get ${\cal P}(t, \omega_{0})= \min\{ n \omega_{0}, 0\}$ with no non-zero root. This suggests that perturbation with $\tilde{\varepsilon_{0}} t$ did not lift the degeneracy and hence, no dispersion can be detected.

In general, when all $\sigma_{i}$s in Eq.~\eqref{eq:Fgen} vanish except the $j$th value, the characteristic polynomial reads
${\cal F}_{\lambda} =  \lambda^{n-j} [ \lambda^{j} + (-1)^{j} \sigma_{j} ]= \lambda^{n-j} {\cal F}'_{\lambda}$ with roots being $\lambda \in \{ 0, [(-1)^{j+1} \sigma_{j}]^{1/j}\}$. As $n-j$ roots are trivial with $\varepsilon=0$, we merely perform tropicalization on ${\cal F}'_{\lambda}$. This characteristic polynomial describes a $j \times j$ matrix whose determinant is $\sigma_{j}$. The characterization of $j$ degenerate eigenvalues for this subsystem is the same as what we have elaborated in the previous paragraph, with tropical roots being $1/j$ for EP$j$s, $1$ for non-defective degeneracies and $0$ when the performed perturbation is incapable of lifting the degeneracy.

The next special case that we cover is a (p,q) torus-knot generically described by~\cite{Hu2022} 
\begin{align}
    H_{(p,q)}=\begin{pmatrix}
        0 & 1 & 0 & \ldots & 0 \\
        0 & 0 & 1 & \ldots & 0 \\
        \vdots & \vdots & \vdots & \ddots & \vdots \\
        0 & 0 & 0 & \ldots & 1 \\
       z^{q} & 0 & 0 & \ldots & 0 
    \end{pmatrix}_{p \times p}.
\end{align}
The characteristic polynomials reads 
\begin{align}
    {\cal F}_{\lambda} = \lambda^{p} + z^{q} =0.
    \label{eq:charpol_pqknot}
\end{align}
The tropicalization of Eq.~\eqref{eq:charpol_pqknot} under linear parametrization of $z = \tilde{z} t$ reads
\begin{align}
    {\cal P}(t, \omega_{0}) = \min\{ p \omega_{0}, q\} ,
\end{align}
where the only nonzero root reads $\omega_{0} = p/q$ and the change of the slope for this root is $p$ indicating that there exists $p$ defective degeneracies with leading order $p/q$. This is in agreement with the solutions of Eq.~\eqref{eq:charpol_pqknot} which are $\lambda_{j} = z^{q/p} e^{2 \pi i (j-1)/p}$ with $z= k_{x} + i k_{y}$ and $j \in \{1, \ldots, p\}$; see also the related discussion on Eq.~\eqref{lidskii-result}. Note that linear perturbation along only $k_{x}$ or $k_{y}$ results in $ {\cal P}(t, \omega_{0}) = \min\{ p \omega_{0}, 0\} ,$ which has no nonzero root~\footnote{This is easy to obtain as $(\tilde{k}_{x}t  + i \tilde{k}_{y})^{q} = \Sigma_{m=0}^{q} \binom{q}{m} (i \tilde{k}_{y})^{q-m} \tilde{k}_{x}^{m} t^{m} $. }. 

To go beyond these particular cases, we now confine ourselves to applying perturbation merely on the $m$th eigenvalue of our system. This leads to having perturbed determinant as $\widetilde{\det[{\cal H}]} = \det[{\cal H}] + \det[{\cal H}]' t$ with $\det[{\cal H}]' = \tilde{\varepsilon}_{m } \prod_{i \neq m} \varepsilon_{i}$ and perturbed traces as $\widetilde{\tr[{\cal H}^{k}]} = \tr[{\cal H}^{k}] + \tr[{\cal H}^{k}]' t$ with $\tr[{\cal H}^{k}]' = \tilde{\varepsilon}_{m} + \sum_{i \neq m} \varepsilon_{i}$. It is straightforward to show that, in terms of $t$, $\sigma_{n}$ is a polynomial of order one and $\sigma_{k}$ with $k \in \{2, \ldots, n-1\}$ are polynomials of order, at most, $k$. For generic cases where all traces and the determinant of $\cal H$ are nonzero, the associated tropical polynomial yields
${\cal P}(\omega_{0}, t) = \min \{ n \omega_{0}, (n-2) \omega_{0}, \ldots, \omega_{0} , 1 \}$ with nonzero roots being $\omega_{0}=1$~\footnote{We note that when all eigenvalues are the same and are perturbed in an identical fashion, $\sigma_{n}$ becomes zero and hence, the tropicalization gives rise to zero tropical roots, i.e., $\omega_{0}=0$.}. This results in linear asymptotic dispersion close to non-defective degeneracies. This conclusion may vary when accidentally or by symmetry, one/multiple traces, or the determinate change~\cite{Delplace2021, Sayyad2022, Sayyad2023, Li2024}.

Let us now comment on the role of sublattice symmetry~(SLS) and pseudo chiral symmetry~(psCS) in determining the asymptotic dispersion. The presence of SLS or psCS enforces $\{ \varepsilon \} = \{-\varepsilon\}$. This ensures vanishing determinants in matrices with odd ranks. For such matrices, $\tr[{\cal H}^{k}]$ is also zero when $k$~(considering $k$ smaller than the rank of the matrix) is odd. These constraints impose that $\sigma_{k}=0$ $\forall k \in \text{odd}$. Hence, the characteristic polynomial $\forall n \in \text{odd}$ casts
\begin{align}
   & {\cal F}_{\lambda} = 
    \lambda^{n}+ \sum\limits_{i \in \text{even}}^{n-1} (-1)^{i}\sigma_{i} \lambda^{n-i} ,
    \end{align}
and for $n \in \text{even}$, it becomes
  \begin{align}  
   {\cal F}_{\lambda} =    \lambda^{n}+ \sum\limits_{i \in \text{even}}^{n-2} (-1)^{i}\sigma_{i} \lambda^{n-i} + (-1)^{n} \sigma_{n}.
\end{align}
Simplifying further would be similar to the previous case as minimum orders after tropicallization is zero unless the model is more specific.

After exploring generic cases, we continue by presenting associated eigensystem characterization of $n \times n$ matrices with $n \leq 4$.

 \section{$2\times2$ matrices}\label{sec:2x2}

 The characteristic polynomial for a generic $2 \times 2$ matrix~($\BH_{2 \times 2}$) is given by~\cite{Sayyad2022}
\begin{align}
    {\cal F}_{\lambda} &=\det[ \lambda \id_{2} - \BH_{2 \times 2}],\nonumber \\
    &= \lambda^2 - \ttr{\BH_{2 \times 2}} \lambda + \det[\BH_{2 \times 2}].
    \label{eq:char2}
\end{align}

The matrix $\BH_{2 \times 2}$ under the similarity transformation $S$ can, in general, cast two possible types of traceless Jordan canonical forms, which are
 \begin{align}
 H_{1,1}& = S \BH_{2 \times 2} S^{-1} = \begin{pmatrix}
         0 & 0\\
         0 & 0
        \end{pmatrix}, \\
  H_{2}& = S \BH_{2 \times 2} S^{-1}= \begin{pmatrix}
           0 & 1\\
           0 & 0
          \end{pmatrix}.
 \end{align}
Here, the subindices in $H_{1,1}$ and $H_{2}$ denote the sizes of underlying Jordan blocks.

Having these Jordan forms, we characterize the corresponding degeneracies of these matrices under generic and non-generic perturbations in the following.

\paragraph{$H_{1,1}$ case.} Since $H_{1,1}$ commutes with any matrix, its most generic traceless perturbation has the form
\begin{equation}
\delta H_{1,1} = \left(
\begin{array}{cc}
 -d_{22} & d_{12} \\
 d_{21} & d_{22} \\
\end{array}
\right), \label{eq:dH11}
\end{equation}
where $d_{ij}\in\mathbb{C}$ with $ij \in \{1,2\}$.
The characteristic polynomial of the full perturbed matrix $H^{\rm tot}_{1,1} =H_{1,1}+\delta H_{1,1}$ is
\begin{equation}
 {\cal F}_{\lambda} = \lambda ^2 -d_{22}^2-d_{12} d_{21},\label{eq:chpol11}
\end{equation}
with the associated roots being
\begin{align}
 \lambda[H^{\rm tot}_{1,1}  ] &= \pm\sqrt{d_{22}^2+d_{12} d_{21}}.
 \label{eq:lamH11}
\end{align}
We note that $\lambda[ H^{\rm tot}_{1,1} ] \neq \lambda[H_{1,1}]=0$.

In general, one may study the leading powers of the eigenvalues under the perturbation using the tropicalization of the characteristic polynomial. To do so, we should fix a specific direction of perturbation such that the perturbation is restricted on a one-dimensional complex linear subspace using $d_{ij} =\tilde{d}_{ij} t$,
where $\tilde{d}_{ij} \in \mathbb{C}$ are constants and $t\in \mathbb{C}$ is the coordinate along the one-dimensional subspace.
With this parametrization, the characteristic polynomial becomes
\begin{equation}
\BF_\lambda(t) = \lambda^2 - (\tilde{d}_{22}^2+\tilde{d}_{12} \tilde{d}_{21})t^2.
\end{equation}
For a general direction of perturbation, for which $(\tilde{d}_{22}^2+\tilde{d}_{12} \tilde{d}_{21})\neq0$, the tropicalization with respect to variable $t$ gives
\begin{equation}
{\cal P}(t,\omega_0) = \min\{2\omega_0, 2\} .
\label{eq:top_h11_w1}
\end{equation}
The only non-zero root of the tropical polynomial is $\omega_0=1$.
This gives us the leading power of the eigenvalue as a function of $t$. The change in the slope at $\omega_{0}=1$. This is in agreement with the asymptotic behavior of eigenvalue in Eq.~\eqref{eq:lamH11} upon substituting $d_{ij} =\tilde{d}_{ij} t$.

In comparison, when $(\tilde{d}_{22}^2+\tilde{d}_{12} \tilde{d}_{21})=0$, the tropicalization of the characteristic polynomial results in
\begin{equation}
{\cal P}(t,\omega_0) = 2\omega_0.
\label{eq:top_h11_w0}
\end{equation}
So, there are no non-zero roots, and the leading power of the eigenvalues is $0$. This agrees with the fact that perturbing the system along this direction does not lift off the degeneracy $\lambda[ H^{\rm tot}_{1,1} ] =\lambda[H_{1,1}]=0$.

\begin{table}
     \centering
     \caption{%Summary of possible nonzero roots of tropical polynomial ${\cal P}(t, \omega_{0})$ for all Jordan forms under the considered perturbation types in $2 \times 2$ matrices.
     Nonzero roots of tropical polynomial ${\cal P}(t, \omega_{0})$ for the generic perturbations of all Jordan forms of $2\times2$ matrices.
     }
     \begin{tabular}{l|l}
     &   $\{$ [ Nonzero $\omega_{0},$ multiplicity] $\}$
     \\
        \hline 
        \hline
      $H^{\rm tot}_{1,1}$ & 
        \{[1,2]\}
        \\
        \hline
       $H^{\rm tot}_{2}$ & 
        $\{$ [$\boldsymbol{\frac{1}{2}}$,2]$\}$
        \\[4pt]
        \hline
\hline
     \end{tabular}
     \vspace{1ex}
     
     {\raggedright \label{tab:rank2} 
      }
\end{table}

\paragraph{$H_{2}$ case. } To find the most general perturbations around $H_{2}$, we first commute a generic $2\times2$ matrix with $H_{2}$, which gives the form of an infinitesimal similarity transformation
\begin{align}
H'_{2 } &= 
(\id_{2} +\delta S) H_{2 }  (\id_{2} +\delta S)^{-1}
,\\
&=H_{2} + [\delta S, H_{2 }] + {\cal O}(\delta S^{2}) ,
\end{align}
with
\begin{align}
 [\delta S, H_{2 }] &= \left[\left(
\begin{array}{cc}
 s_{11} & s_{12} \\
 s_{21} & s_{22} \\
\end{array}
\right), \begin{pmatrix}
           0 & 1\\
           0 & 0
          \end{pmatrix}
          \right],\\
          &= \left(
\begin{array}{cc}
 -s_{21} & s_{11}-s_{22} \\
 0 & s_{21} \\
\end{array}
\right).
\label{eq:trivH2}
\end{align}
As any similarity transformation keeps the eigenvalues invariant, eigenvalues of perturbed $H'_{2 }$ are identical to eigenvalues of $H_{2 }$ which is $\lambda[H'_{2}]= \lambda[H_{2 }] =0$. For this reason, the perturbation $ [\delta S, H_{2 }] $ is known as a trivial perturbation.
The nontrivial perturbation can be obtained by comparing Eq.~\eqref{eq:dH11} and Eq.\eqref{eq:trivH2}. To be more precise, after quotienting out infinitesimal similarity transformations from Eq.~\eqref{eq:dH11}, we obtain the nontrivial traceless perturbation given by
\begin{equation}
 \delta H_{2} = \begin{pmatrix}
                 0 & 0\\
                 d_{21} & 0
                \end{pmatrix},\, \, H^{\rm tot}_{2} =H_{2}+\delta H_{2} = \begin{pmatrix}
                0 & 1\\
                d_{21} & 0
                \end{pmatrix},
                \label{eq:dH2}
\end{equation}
where $d_{21} \in \mathbb{C}$. The characteristic polynomial and the eigenvalues of the perturbed matrix are then
\begin{align}
 {\cal F}_{\lambda}  &= \lambda^2 - d_{21},
\\
 \lambda[H^{\rm tot}_{2} ] &= \pm\sqrt{d_{21}}.
\end{align}
We note that the eigenvalues of the perturbed Hamiltonian is no longer degenerate as $\lambda[H^{\rm tot}_{2} ] \neq \lambda[H_{2}] $ for nonzero $d_{21}$.

The tropical polynomial for parametrized variable $d_{21} =\tilde{d}_{21} t$ based on the characteristic polynomial reads
\begin{align}
    {\cal P}(t, \omega_{0}) = \min\{2 \omega_{0},1\}, \label{eq:top_h2_w12}
\end{align}
with the root $\omega_{0} =1/2$. Again, the change of slope at the root is $2$, which translates to $2$ eigenvalues with leading square-root behavior, i.e. the multiplicity of eigenvalues is $2$. This suggests that varying $d_{21}$ displays EP2s with second-root asymptotic dispersion.

We summarize emerging tropical roots and their multiplicities for generic perturbations of Jordan forms of
$2 \times 2$ matrices in Table~\ref{tab:rank2}.

\begin{table}
     \centering
     \caption{Summary of possible roots of tropical polynomial ${\cal P}(t, \omega_{0})$ for all Jordan forms under considered perturbations in $3 \times 3$ matrices. Results for generic perturbations are highlighted in bold.}
     \begin{tabular}{l|l}
     &  $\{$ [ Nonzero $\omega_{0},$ multiplicity] $\}$
     \\
        \hline 
        \hline
      $H^{\rm tot}_{1,1,1}$ & 
        \{ {\bf [1,3]}, [1,2] \}
        \\[4pt]
        \hline
       $H^{\rm tot}_{2,1}$ & 
        $\{$ {\bf[($\boldsymbol{\frac{1}{2}}$, 2) , (1,1)]}, [$\frac{2}{3}$,3], [1,2], [$\frac{1}{2}$, 2]$\}$
           \\[4pt]
        \hline
       $H^{\rm tot}_{3}$ & 
        $\{$ {\bf[$\frac{1}{3}$, 3]}, [$\frac{1}{2}$, 2] $\}$
        \\[4pt]
        \hline
\hline
     \end{tabular}
     \vspace{1ex}
     
     {\raggedright \label{tab:rank3} 
     
      For cases where the total multiplicity is smaller than the rank of the matrix, there exists zero eigenvalues in the spectrum. When the system hosts multiple roots, each root with its associated multiplicity appear together as (tropical root, multiplicity)
      }
\end{table}

\section{$3 \times 3$ matrices}\label{sec:3x3}
The characteristic polynomial for a generic $3 \times 3$ matrix~($\BH_{3 \times 3}$) reads~\cite{Sayyad2022}
\begin{align}
    {\cal F}_{\lambda} = \lambda^3 - a_{3} \lambda^{2} + b_{3} \lambda -c_{3}=0,
    \label{eq:char3}
\end{align}
where 
\begin{align}
    a_{3} &=\ttr{\BH_{3 \times 3}},\\
    b_{3} &= \frac{(\ttr{\BH_{3 \times 3}})^2 - \ttr{\BH_{3 \times 3}^2}}{2} ,\\
    c_{3} &=\det[\BH_{3 \times 3}].
\end{align}

In this case, three different types of Jordan block structures may occur, which are
\begin{align}
 H_{1,1,1} &= \begin{pmatrix}
              0 & 0 & 0\\
              0 & 0 & 0\\
              0 & 0 & 0
             \end{pmatrix},\\
 H_{2,1} &= \begin{pmatrix}
            0 & 1 & 0\\
            0 & 0 & 0\\
            0 & 0 & 0
           \end{pmatrix},\\
 H_{3} &= \begin{pmatrix}
           0 & 1 & 0\\
           0 & 0 & 1\\
           0 & 0 & 0
         \end{pmatrix}.
\end{align}

In the following, we present how asymptotic dispersion relations of degeneracies in these systems vary when we apply generic and non-generic perturbations.

\paragraph{$H_{1,1,1}$ case. }
As $H_{1,1,1}$ commutes with any $3\times3$ matrix, $\delta H_{1,1,1}$ is an arbitrary traceless $3\times3$ which reads
\begin{align}
    H^{\rm tot}_{1,1,1} =
   \left(
\begin{array}{ccc}
 d_{11} & d_{12} & d_{13} \\
 d_{21} & -d_{11}-d_{33} & d_{23} \\
 d_{31} & d_{32} & d_{33} \\
\end{array}
\right),\label{eq:Htot111}
\end{align}
where $H^{\rm tot}_{1,1,1} = H_{1,1,1} +\delta H_{1,1,1}$ and $d_{ij} \in \mathbb{C}$ with $i,j \in \{1,2,3\}$.
The characteristic polynomial of perturbed Hamiltonian is a depressed cubic,
\begin{equation}
{\cal F}_{\lambda} = \lambda^3+p\lambda+q,
\label{eq:charpolH111}
\end{equation}
with $p$ and $q$ being
\begin{align}
   - p =& d_{11}^2+d_{33} d_{11}+d_{33}^2 \nonumber
   \\ &+d_{12} d_{21}+d_{13} d_{31}+d_{23} d_{32}
    \label{eq:pH111}
    ,\\
    -q =& d_{13} d_{31} d_{33}-d_{33}^2 d_{11}+d_{13} d_{31} d_{11} \nonumber \\
    &+d_{12} d_{23} d_{31}+d_{13} d_{21} d_{32}-d_{33} d_{11}^2\nonumber \\
   &-d_{12} d_{21} d_{33} -d_{23} d_{32} d_{11}.
      \label{eq:qH111}
\end{align}

The three roots of $\BF_{\lambda}$ are~\cite{Kurosh-1972}
\begin{align}
\lambda_1 & = \alpha +\beta,\nonumber\\
\lambda_2 & = e^{2i\pi/3}\alpha+e^{-2i\pi/3}\beta,\label{eq:eigvals3}\\
\lambda_3 & = e^{-2i\pi/3}\alpha+e^{2i\pi/3}\beta,\nonumber
\end{align}
where
\begin{align}
\alpha &= \sqrt[3]{-q + \sqrt{p^3+q^2}},\label{eq:alphadef}
 \\
\beta &= -p/\alpha = -\sqrt[3]{q+\sqrt{p^3+q^2}}.
\label{eq:betadef}
\end{align}
Here, it is assumed that the same branches of the complex root function are chosen in Eqs.~\eqref{eq:alphadef} and~\eqref{eq:betadef}, so that $\beta=-p/\alpha$ is satisfied.

Notice that $p$ in Eq.~\eqref{eq:pH111} is a homogeneous polynomial of order $2$ in the coefficients of $\delta H_{1,1,1}$, and $q$ in Eq.~\eqref{eq:qH111} is a homogeneous polynomial of order $3$ in the same coefficients. $p^3+q^2$ is also a homogeneous polynomial of order $6$. Now, imagine we traverse an infinitesimal loop around $p=q=0$ such that the expressions $\alpha$ and $\beta$ do not vanish. Then, $\sqrt{p^3+q^2}$ and  $q$ acquire a phase of $6 \pi$ so that after taking the third roots, there remains a phase of $2\pi$, i.e., there is no braiding.

As in the case of $2 \times 2$ matrices, it is convenient to choose a one-dimensional complex linear subspace by parametrizing entries of $H_{1,1}^{\rm tot}$ by $d_{ij} = \tilde{d}_{ij} t$ where again $t$ being the coordinate along this one-dimensional subspace. The characteristic polynomial is restricted to the subspace as
\begin{equation}
\BF_\lambda (t)= \lambda^3 + (\tilde p t^2) \lambda + \tilde q t^3.
\end{equation}
Again, we consider a general direction where $\tilde p,\tilde q\neq 0$. Here, the tropicalization of the characteristic polynomial reads
\begin{equation}
\BP(t,\omega_0) = \min\{3\omega_0,2+\omega_0,3\} .
\label{eq:top_h111_w1}
\end{equation}
There is a single tropical root $\omega_0 = 1$. The change of slope~(multiplicity) at this point is $3$. Therefore, the tropical polynomial predicts three eigenvalues with leading power $1$, expected for non-defective degeneracies.

The tropical root remains unchanged upon having $\tilde p=0,\tilde q\neq0$. However, in the case where $\tilde p\neq0, \tilde q=0$, the tropicalization of characteristic polynomial gives
\begin{equation}
\BP(t,\omega_0) = \min\{3\omega_0,2+\omega_0\}.
\end{equation}
The only non-zero root is still $1$; however, the multiplicity at the root is now only $2$. Thus, we have two eigenvalues which disperse linearly in $t$. The third eigenvalue is $\lambda=0$, a flat degeneracy that remains unchanged upon perturbing the system with $\tilde p\neq0, \tilde q=0$.

Finally, if both $\tilde p$ and $\tilde q$ are zero, the tropicalization is trivial $\min\{3\omega_0\}$, and there are no non-zero tropical roots, i.e., all eigenvalues have leading power zero.

\paragraph{$H_{2,1}$ case. }
For $H_{2,1}$, the infinitesimal similarity transformation $\delta S$ results in a trivial perturbation $[\delta S, H_{2,1 }]$ which reads
\begin{align}
H'_{2,1 } &= 
(\id_{2} +\delta S) H_{2,1 }  (\id_{2} +\delta S)^{-1}
,\nonumber \\
&=H_{2,1} + [\delta S, H_{2,1 }] + {\cal O}(\delta S^{2})
,\\
 [\delta S, H_{2,1 }] &= 
 \left[\left(
\begin{array}{ccc}
 s_{11} & s_{12} & s_{13} \\
s_{21} & s_{22} & s_{23} \\
 s_{31} & s_{32} & s_{33} \\
\end{array}
\right),
       \begin{pmatrix}
           0 & 1 & 0\\
           0 & 0 & 0\\
           0 & 0 & 0
         \end{pmatrix}
\right] , \nonumber \\
&=
\left(
\begin{array}{ccc}
 -s_{21} & s_{11}-s_{22} & -s_{23} \\
 0 & s_{21} & 0 \\
 0 & s_{31} & 0 \\
\end{array}
\right). \label{eq:trivH21}
\end{align}
We emphasize that $\lambda[H'_{2,1}] = \lambda[H_{2,1}]$.
Quotienting out the r.h.s of Eq.~\eqref{eq:trivH21} from traceless matrices, we find the nontrivial traceless perturbations, e.g., given in Eq.~\eqref{eq:Htot111} which reads
\begin{align}
 \delta H_{2,1} &= \left(
\begin{array}{ccc}
 0 & 0 & 0 \\
 d_{21} & -d_{33} & d_{23} \\
 d_{31} & 0 & d_{33} \\
\end{array}
\right),
\\
 H^{\rm tot}_{2,1} &=\left(
\begin{array}{ccc}
 0 & 1 & 0 \\
 d_{21} & -d_{33} & d_{23} \\
 d_{31} & 0 & d_{33} \\
\end{array}
\right),
\end{align}
where $H^{\rm tot}_{2,1} = H_{2,1} + \delta H_{2,1} $.
The corresponding characteristic polynomial is 
\begin{equation}
 {\cal F}_{\lambda} = \lambda ^3 - \left(d_{33}^2+d_{21}\right) \lambda -d_{23} d_{31}+d_{21} d_{33},\label{chpol21}
\end{equation}
where $p=-(d_{33}^2+d_{21})$  and $q=d_{21} d_{33}-d_{23} d_{31}$ and roots of ${\cal F}_{\lambda}$ are given in Eq.~\eqref{eq:eigvals3}.

Now, the coefficient $q$ is homogeneous of the second order. However, the coefficient $p$ is not a homogeneous polynomial, but under the assumption that all the parameters are similar in magnitude and small, we may neglect $d_{33}^2$ in comparison with $d_{21}$ and then attribute linear order to it. Using the power counting arguments, we can then neglect $q^2$ in comparison with $p^3$ and $q$ in comparison with $\sqrt{p^3}$.
Writing $u=|u|e^{i\varphi}$, we can approximate the eigenvalues of $H_{2,1}^{\rm tot}$ as
\begin{equation}
 \lambda= 2i\sqrt{|u|}\sin{\left[\frac{\varphi}{2} + \frac{2\pi}{3}k\right]},\quad k \in \{0,1,2\},
\end{equation}
 using 
 \begin{equation}
 \alpha = \sqrt{|u|}e^{i\varphi/2},\, \, \beta = -u/\alpha = -\sqrt{|u|}e^{-i\varphi/2}.
\end{equation}
Traversing an infinitesimal loop without crossing degeneracies, keeps $\lambda_1$ intact and exchanges $\lambda_{2}$ and $\lambda_{3}$. 
Therefore, we observe a Hopf link (and a separate loop), which is the typical braid for second-order EPs~\cite{Chen2024}.

\begin{figure}
    \centering
 \includegraphics[width=0.5\textwidth]{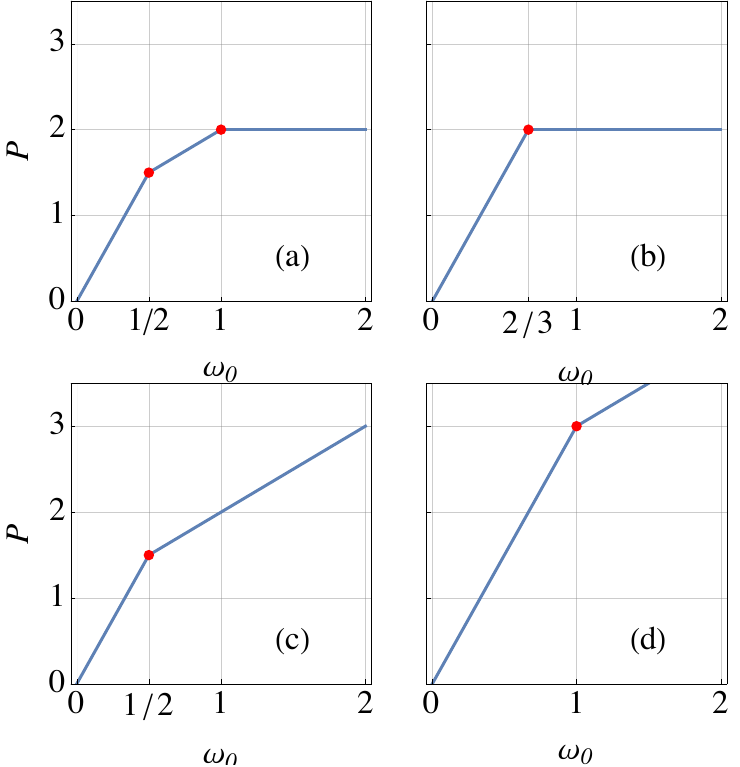}
    \caption{Possible tropical polynomials in $H_{2,1,1}$ with explicit forms given in Eq.~\eqref{eq:trop_poly_h21a}~(a), Eq.~\eqref{eq:trop_poly_h21b}~(b), Eq.~\eqref{eq:trop_poly_h21c}~(c),  and Eq.~\eqref{eq:trop_poly_h21d}~(d).
    Red points indicate the roots of the tropical polynomials.
    }
    \label{fig:trop_poly_h21}
\end{figure}

Reparametrizing matrix elements as $d_{ij} =\tilde{d}_{ij} t$, we rewrite the characteristic polynomial~\eqref{chpol21} as
\begin{equation}
\BF_\lambda (t)= \lambda^3 - (\tilde d_{33}^2 t^2 + \tilde d_{21} t)\lambda - \tilde q_{2,1}t^2,
\end{equation}
where
\begin{equation}
\tilde q_{2,1} = \tilde d_{23}\tilde d_{31}-\tilde d_{21}\tilde d_{33}.
\end{equation}
For a general direction of perturbation, $\tilde d_{21},\tilde q_{2,1}\neq0$, the tropicalization of the characteristic polynomial shown in Fig.~\ref{fig:trop_poly_h21}~(a) is
\begin{equation}
\BP(t,\omega_0) = \min\{3\omega_0,1+\omega_0,2\}.
\label{eq:trop_poly_h21a}
\end{equation}
Two non-zero tropical roots~(red circle points) are $\omega_0=1/2$ and $\omega_0=1$ with multiplicities $2$ and $1$, respectively. As expected, this implies a doublet of eigenvalues with a square-root leading behavior and one eigenvalue with a linear asymptotic dispersion relation.

When $\tilde{d}_{21}=0$, $\tilde{d}_{33}\neq 0$ and $\tilde q_{2,1}\neq 0$, the tropical polynomial plotted in Fig.~\ref{fig:trop_poly_h21}~(b) reads
\begin{equation}
\BP(t,\omega_0) = \min\{3\omega_0, 2+\omega_0, 2\}.
\label{eq:trop_poly_h21b}
\end{equation}
Here, the tropical root is $2/3$~(red circle point) with multiplicity $3$.
Interestingly, the splitting of eigenvalues resembles the generic $H_3$ case~(see below) albeit with non-generic exponent.
 Considering $\tilde d_{21}=0$, $\tilde d_{33}\neq0$, $\tilde q_{2,1}=0$ changes the tropical polynomial displayed in Fig.~\ref{fig:trop_poly_h21}~(c) to
\begin{equation}
\BP(t,\omega_0) = \min\{3\omega_0,2+\omega_0\}.
\label{eq:trop_poly_h21d}
\end{equation}
The tropical root is then $\omega_0=1$~(red circle point) with multiplicity $2$. The third eigenvalue $\lambda=0$ has the leading power $0$. This means that there exist two degeneracies with linear dispersion in $t$ and a flat third eigenvalue at $\lambda=0$. When $\tilde d_{21}\neq0$, $\tilde q_{2,1}= 0$, the tropical polynomial shown in Fig.~\ref{fig:trop_poly_h21}~(d) becomes
\begin{equation}
\BP(t,\omega_0) = \min\{3\omega_0, 1+\omega_0\},
\label{eq:trop_poly_h21c}
\end{equation}
with the tropical root being $1/2$~(red circle point) with multiplicity $2$. Since there are no more non-zero tropical roots, the third eigenvalue is $\lambda=0$ with leading power $0$. Lastly, when all variables vanish, the trivial tropicalization reads $\BP(t, \omega_{0}) = \min \{3\omega_0 \}$, and subsequently, all eigenvalues have leading power $0$. In other words, the imposed perturbation does not lift degeneracies of $H_{2,1}$.

\paragraph{$H_{3}$ case. }
For $H_{3}$, the infinitesimal similarity transformation $\delta S$ results in a trivial perturbation $[\delta S, H_{3}]$ which reads~\cite{jiang2020thesis}
\begin{align}
[\delta S, H_{3}] &=
 \left[\left(
\begin{array}{ccc}
 s_{11} & s_{12} & s_{13} \\
 s_{21} & s_{22} & s_{23} \\
 s_{31} & s_{32} & s_{33} \\
\end{array}
\right),
       \begin{pmatrix}
           0 & 1 & 0\\
           0 & 0 & 1\\
           0 & 0 & 0
         \end{pmatrix}
\right] ,\nonumber\\
&= \left(
\begin{array}{ccc}
 -s_{21} & s_{11}-s_{22} & s_{12}-s_{23} \\
 -s_{31} & s_{21}-s_{32} & s_{22}-s_{33} \\
 0 & s_{31} & s_{32} \\
\end{array}
\right). \label{eq:trivH3}
\end{align}
Quotienting out the r.h.s of Eq.~\eqref{eq:trivH3} from traceless matrices in Eq.~\eqref{eq:Htot111}, we find the generic nontrivial perturbation which is
\begin{equation}
 \delta H_3 = \begin{pmatrix}
               0 & 0 & 0\\
               0 & 0 & 0\\
               d_{31} & d_{32} & 0
              \end{pmatrix},\,
H^{\rm tot}_{3} = \begin{pmatrix}
               0 & 1 & 0\\
               0 & 0 & 1\\
              d_{31} & d_{32} & 0
              \end{pmatrix},
              \label{eq:Htot3}
\end{equation}
where $ H^{\rm tot}_{3} = H_3 + \delta H_3$.

The associated characteristic polynomial then reads
\begin{align}
    {\cal F}_{\lambda} &= \lambda ^3 -d_{32} \lambda -d_{31},\label{eq:chpol3}
\end{align}
 where $\lambda[H^{\rm tot}_{3} ] \neq \lambda[H_{3} ]$.

Tracking the eigenvalue braiding along an infinitesimal loop, not crossing any degeneracies, gives the standard trefoil knot~\cite{Patil2022, Chen2024}.

Introducing the reparametrization $d_{ij} = \tilde{d}_{ij} t$ with $t$ being along a fixed one-dimensional complex linear subspace, we rewrite the characteristic polynomial in Eq.~\eqref{eq:chpol3} as
\begin{equation}
\BF_\lambda(t) = \lambda^3 - (\tilde{d}_{32} t)\lambda - \tilde d_{31} t.
\end{equation}
For the general direction corresponding to $\tilde d_{32},\tilde d_{31}\neq0$, 
the tropicalization gives
\begin{equation}
\BP(t,\omega_0) = \min\{3\omega_0, \omega_{0}+1,1\}.
\end{equation}
This polynomial has a single root $\omega_0=1/3$ with multiplicity $3$~(based on the slope change).
This root remain unchanged when $\tilde{d}_{32}=0$. This is because there exists a similarity matrix $S'$ that can transform $H^{\rm tot}_{3}$ with nonzero $d_{ij}$ to $H^{\rm tot}_{3}$ with only $d_{31} \neq 0$~\cite{Simtrans_Jn}.

However, setting $d_{31}=0$ modifies the tropical polynomial into
\begin{equation}
\BP(t,\omega_0) = \min\{3\omega_0, 1+\omega\}.
\end{equation}
Then, the single non-zero root is $\omega_0=1/2$ with multiplicity $2$. Since we have three eigenvalues, the third would have the leading power $0$, corresponding to non-dispersive $\lambda=0$.

We summarized all nonzero tropical roots and their associated multiplicities for different Jordan forms of $3 \times 3$ matrices under generic perturbations in Table.~\ref{tab:rank3}.

\section{$4 \times 4$ matrices}\label{sec:4x4}

The characteristic polynomial for a generic $4 \times 4$ matrix~($H_{4 \times 4}$) is given by~\cite{Sayyad2022}
\begin{align}
    {\cal F}_{\lambda} = \lambda^4 - a_{4} \lambda^{3} + b_{4} \lambda^{2} -c_{4} \lambda +d_{4} =0,
    \label{eq:char4}
\end{align}
where 
\begin{align}
    a_{4} &=\ttr{H_{4 \times 4}},\\
    b_{4} &= \frac{(\ttr{H_{4 \times 4}})^2 - \ttr{H_{4 \times 4}^2}}{2} ,\\
    c_{4} &= \frac{(\ttr{H_{4 \times 4}})^3 - 3 \ttr{H_{4 \times 4}} \ttr{H_{4 \times 4}^2} + 2 \ttr{H_{4 \times 4}^{3}}}{6},\\
    d_{4} &=\det[H_{4 \times 4}].
\end{align}

Similarity transformations on $H_{4 \times 4}$ matrices result in obtaining one of the following five Jordan blocks
\begin{align}
    H_{1,1,1,1}&=\begin{pmatrix}
        0 & 0 & 0 & 0 \\
        0 & 0 & 0 & 0 \\
        0 & 0 & 0 & 0 \\
        0 & 0 & 0 & 0 
    \end{pmatrix},
    \,
     H_{2,1,1}=\begin{pmatrix}
        0 & 1 & 0 & 0 \\
        0 & 0 & 0 & 0 \\
        0 & 0 & 0 & 0 \\
        0 & 0 & 0 & 0 
    \end{pmatrix},
    \\
    H_{2,2}&=\begin{pmatrix}
        0 & 1 & 0 & 0 \\
        0 & 0 & 0 & 0 \\
        0 & 0 & 0 & 1 \\
        0 & 0 & 0 & 0 
    \end{pmatrix},
    \\
     H_{3,1}&=\begin{pmatrix}
        0 & 1 & 0 & 0 \\
        0 & 0 & 1 & 0 \\
        0 & 0 & 0 & 0 \\
        0 & 0 & 0 & 0 
    \end{pmatrix},
    \,
     H_{4}=\begin{pmatrix}
        0 & 1 & 0 & 0 \\
        0 & 0 & 1 & 0 \\
        0 & 0 & 0 & 1 \\
        0 & 0 & 0 & 0 
    \end{pmatrix}.
\end{align}

We now proceed with obtaining leading orders of degeneracies in these systems under generic and non-generic perturbations.

\begin{table*}
     \centering
     \caption{Summary of possible roots of tropical polynomial ${\cal P}(t, \omega_{0})$ for all Jordan forms under considered perturbations in $4 \times 4$ matrices. Results for generic perturbations are highlighted in bold.}
     \begin{tabular}{l|l}
     &  $\{$ [ Nonzero $\omega_{0},$ multiplicity] $\}$
     \\
        \hline 
        \hline
      $H^{\rm tot}_{1,1,1,1}$ & 
        \{  {\bf [1,4]}, [1,3], [1,2]\}
        \\
        \hline
       $H^{\rm tot}_{2,1,1}$ & 
        $\{$ {\bf [($\boldsymbol{\frac{1}{2}}$, 2) , (1,2)]}, [($\frac{2}{3}$,3),(1,1)], [($\frac{1}{2}$, 2) , (1,1)]$\}$
        \\[3pt]
        \hline
       $H^{\rm tot}_{2,2}$ & 
        $\{$ {\bf [$\boldsymbol{\frac{1}{2}}$,4]}, [($\frac{1}{2}$,2),(1,1)], [($\frac{2}{3}$,3),(1,1)]$\}$
           \\[3pt]
        \hline
       $H^{\rm tot}_{3,1}$ & 
        $\{${\bf [($\boldsymbol{\frac{1}{3}}$,3),(1,1)]}, [($\frac{1}{2}$,2),(1,1)], [($\frac{1}{2}$,4)], [($\frac{1}{3}$,3),(1,1)]$\}$
         \\[3pt]
        \hline
       $H^{\rm tot}_{4}$ & 
        $\{${\bf [$\frac{1}{4}$,4]}, [$\frac{1}{3}$,3], [$\frac{1}{2}$,2]$\}$
        \\[3pt]
        \hline
\hline
     \end{tabular}
     \vspace{1ex}
     
     {\raggedright \label{tab:rank4} 
     
      For cases where the total multiplicity is smaller than the rank of the matrix, there exists zero eigenvalues in the spectrum. When the system hosts multiple roots, each root with its associated multiplicity appear together as (tropical root, multiplicity).
      }
\end{table*}

\paragraph{$H_{1,1,1,1}$ case. }

Similar to previous cases, i.e. $H^{\rm tot}_{1,1}, H^{\rm tot}_{1,1,1}$, the generic traceless perturbed form of $H_{1,1,1,1}$ reads
\begin{align}
H^{\rm tot}_{1,1,1,1}=&
\left(
\begin{array}{cccc}
 d_{11} & d_{12} & d_{13} & d_{14} \\
 d_{21} & d_{22}-d_{11} & d_{23} & d_{24} \\
 d_{31} & d_{32} & -d_{44}-d_{22} & d_{34} \\
 d_{41} & d_{42} & d_{43} & d_{44} \\
\end{array}
\right), \label{eq:dH1111}
\end{align}
where $d_{ij} \in \mathbb{C}$ with $i,j \in \{1,2,3,4\}$.
The associated characteristic polynomial is lengthy, and we refrain from presenting that here. However, it takes the general form
\begin{equation}
\BF_\lambda = \lambda^4 + r \lambda^2 - p\lambda + q,\label{eq:chpol-gen4}
\end{equation}
where $r$, $p$ and $q$ are homogenous polynomials of order $2$, $3$, and $4$ in terms of $d_{ij}$. 

Employing the parametrization $d_{ij}=\tilde{d}_{ij} t$ enables us to rewrite ${\cal F}_{\lambda}$ as
\begin{equation}
\BF_\lambda(t) = \lambda^4 +(\tilde{r} t^2) \lambda^2 - (\tilde{p} t^3) \lambda + \tilde{q} t^4.
\end{equation}
The corresponding tropicalization with respect to $t$ then reads
\begin{align}
\BP(t,\omega_0) &= \min\{4\omega_0, 2+2\omega_0, 3+\omega_0, 4\}.
\end{align}
There is a single tropical root $\omega_0=1$ with multiplicity $4$, which is typical for non-defective degeneracies.
When $\tilde{r} =0$, the characteristic polynomial becomes
    \begin{align}
        {\cal F}_{\lambda} (t)
        = \lambda^{4} - (\tilde{p} t^{3}) \lambda + \tilde{q} t^{4} ,
    \end{align}
    associated with the tropical polynomial of 
    \begin{align}
        {\cal P} (t, \omega_{0}) = \min \{ 4 \omega_{0}, \omega_{0}+3, 4\}.
    \end{align}
The nonzero root of this polynomial is then $\omega_{0}=1$ with multiplicity $3$. Note that the fourth eigenvalue is $\lambda=0$, which remains unperturbed under the perturbation with $\tilde{r} =0$. Considering $\tilde{p}=0$, the characteristic polynomial then becomes
\begin{align}
        {\cal F}_{\lambda}(t) 
        = \lambda^{4} - (\tilde{r} t^{2}) \lambda^{2} + \tilde{q} t^{4} .
    \end{align}
    The tropicalization then results in 
    \begin{align}
        {\cal P} (t, \omega_{0}) = \min \{ 4 \omega_{0}, 2\omega_{0}+2, 4\}.
    \end{align}
    This polynomial has a single tropical root $\omega_{0}=1$ with multiplicity $2$. In addition to these two linearly dispersive degeneracies, there exist two other eigenvalues $\lambda=0$, which did not change under the imposed perturbation. 
When $\tilde{q}=0$, the characteristic polynomial casts
    \begin{equation}
\BF_\lambda(t) = \lambda^4 +(\tilde r t^2) \lambda^2 - (\tilde p t^3) \lambda ,
\end{equation}
with the associated tropical polynomial being
 \begin{align}
        {\cal P} (t, \omega_{0}) = \min \{ 4 \omega_{0}, 2\omega_{0}+2, \omega_{0} +3 \}.
    \end{align}
    Here, the tropical root is $\omega_{0}=1$ with multiplicity $2$. Similar to the perturbation with $\tilde{p}=0$, we have two unchanged eigenvalues $\lambda=0$ and two eigenvalues with linear asymptotic dispersion relations.

\paragraph{$H_{2,1,1}$ case. }

Applying an infinitesimal similarity transformation with $(\id_{4} + \delta S )$ on $H_{2,1,1}$ results in obtaining trivial perturbations~\footnote{The procedure is similar to the previous cases for $3 \times 3$ Jordan forms; see also Ref.~\cite{Jiang2020}.}.

\begin{figure}
    \centering
 \includegraphics[width=0.49\textwidth]{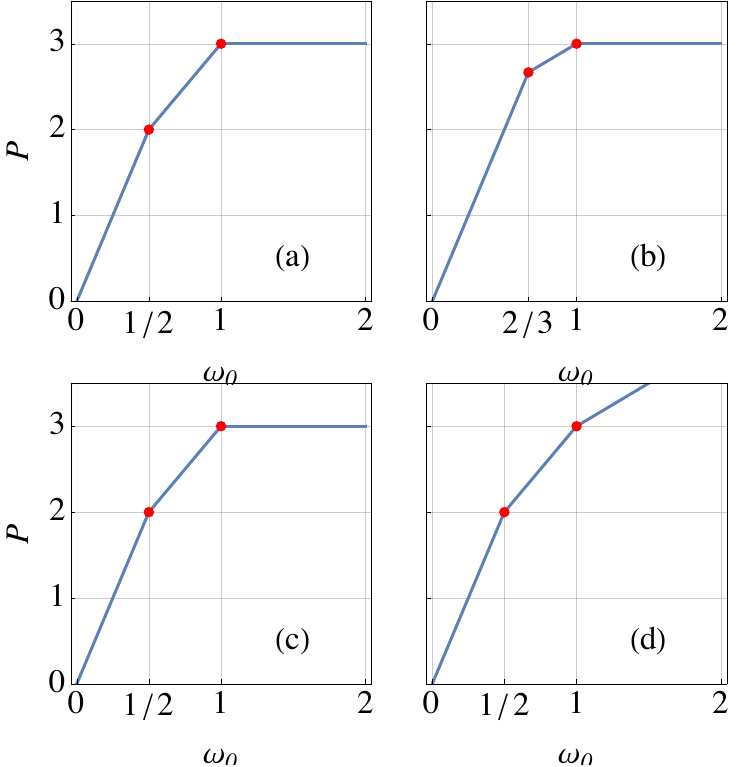}
    \caption{Possible tropical polynomials in $H_{2,1,1}$ with explicit forms given in Eq.~\eqref{eq:trop_polyh211a}~(a), Eq.~\eqref{eq:trop_polyh211b}~(b), Eq.~\eqref{eq:trop_polyh211c}~(c),  and Eq.~\eqref{eq:trop_polyh211d}~(d).
    Red points indicate the roots of the tropical polynomials.
    }
    \label{fig:trop_poly_h211}
\end{figure}

Having these trivial contributions, we realize that the generic nontrivial traceless perturbation $\delta H_{2,1,1}$ should obey $(\delta H_{2,1,1})_{41} \neq  0$ , $(\delta H_{2,1,1})_{43} \neq  0$, $(\delta H_{2,1,1})_{44} \neq  0$, $(\delta H_{2,1,1})_{31} \neq  0$ , $(\delta H_{2,1,1})_{33} \neq  0$, $(\delta H_{2,1,1})_{34} \neq  0 $ , $(\delta H_{2,1,1})_{21} \neq  0 $ $(\delta H_{2,1,1})_{23} \neq  0$, and
$(\delta H_{2,1,2} )_{24} \neq 0 $ such that
\begin{align}
     H^{\rm tot}_{2,1,1} = H_{2,1,1} +\delta H_{2,1,1} =
\left(
\begin{array}{cccc}
 0 & 1 & 0 & 0 \\
 d_{21} & 0 & d_{23} & d_{24} \\
 d_{31} & 0 & -d_{44} & d_{34} \\
 d_{41} & 0 & d_{43} & d_{44} \\
\end{array}
\right).
\end{align}
The associated characteristic polynomial reads
\begin{align}
  {\cal F}_{\lambda} &= \lambda^4 -\left(d_{44}^2+d_{21}+d_{34} d_{43}\right) \lambda ^2
  \nonumber \\
  &
  -\left(d_{23} d_{31} +d_{24} d_{41}\right) \lambda
  \nonumber \\
  &
 +d_{21} d_{44}^2+d_{23} d_{31} d_{44}-d_{24} d_{41} d_{44}
  \nonumber \\
  &
  -d_{21} d_{34} d_{43}-d_{24} d_{31} d_{43}
  -d_{23} d_{34} d_{41}
  .\label{eq:chpol211}
\end{align}
Here, the coefficients of the linear and the free terms are homogeneous polynomials of orders $2$ and $3$ in the perturbation parameters, respectively. Therefore, after parametrization $d_{ij}=\tilde{d}_{ij} t$, we obtain the characteristic polynomial as
\begin{align}
\BF_\lambda(t) = \lambda^4 &- (\tilde d_{21} t + \tilde d_{44}^2 t^2 + \tilde d_{34}\tilde d_{43}t^2)\lambda^2 \\& - (\tilde p_{2,1,1} t^2)\lambda + (\tilde q_{2,1,1} t^3),
\end{align}
where $\tilde{p}_{2,1,1}= \tilde{d}_{23} \tilde{d}_{31} +\tilde{d}_{24} \tilde{d}_{41}$ and $\tilde{q}_{2,1,1} = \tilde{d}_{21} \tilde{d}_{44}^2+\tilde{d}_{23} \tilde{d}_{31} \tilde{d}_{44}-\tilde{d}_{24} \tilde{d}_{41} \tilde{d}_{44}
  -\tilde{d}_{21} \tilde{d}_{34} \tilde{d}_{43}-\tilde{d}_{24} \tilde{d}_{31} \tilde{d}_{43}
  -\tilde{d}_{23} \tilde{d}_{34} \tilde{d}_{41}$.
In the general case, where $\tilde d_{21},\tilde p_{2,1,1},\tilde q_{2,1,1}\neq0$,
the tropicalization leads to
\begin{equation}
\BP(t,\omega_0) = \min\{4\omega_0, 1+2\omega_0, 2+\omega_0, 3\},
\label{eq:trop_polyh211a}
\end{equation}
which is shown in Fig.~\ref{fig:trop_poly_h211}~(b).
This gives us two roots, which are $\omega_{0}=1/2$ and $\omega_{0}=1$~(red points) with multiplicities $2$ and $2$, respectively. This indicates a doublet of square-root eigenvalues~(EP2) and two linear eigenvalues. 

When $\tilde{d}_{21}=0$, the characteristic polynomials becomes
    \begin{align}
\BF_\lambda(t) = \lambda^4 &- ( \tilde d_{44}^2  + \tilde d_{34}\tilde d_{43}) t^2\lambda^2  \nonumber
\\& - (\tilde p_{2,1,1} t^2)\lambda + (\tilde q_{2,1,1} t^3).
\end{align}
The tropicalization then gives
\begin{align}
    {\cal P}(t, \omega_{0}) = \min\{
    4 \omega_{0} , 2 \omega_{0}+2, \omega_{0}+2, 3
    \},
    \label{eq:trop_polyh211b}
\end{align}
with tropical roots being $\omega_{0}=1$ with multiplicity $1$ and $\omega_{0}=2/3$ with multiplicity $3$. Red symbols in Fig.~\ref{fig:trop_poly_h211}~(b) indicate these roots. 

Upon having $\tilde{p}_{2,1,1}=0$, the characteristic polynomial and the tropical polynomial read
\begin{align}
\BF_\lambda (t) =& \lambda^4 - (\tilde d_{21} t + \tilde d_{44}^2 t^2 + \tilde d_{34}\tilde d_{43}t^2)\lambda^2 \nonumber
\\& + (\tilde q_{2,1,1} t^3),
\\
{\cal P}(t, \omega_{0} ) =&
\min\{ 4 \omega_{0}, 2 \omega_{0}+1 , 3\}.
\label{eq:trop_polyh211c}
 \end{align}
 The roots of ${\cal P}$ are $1/2$ with multiplicity $2$ and $1$ with multiplicity $2$. Figure~\ref{fig:trop_poly_h211}~(c) presents the tropical polynomial in Eq.~\eqref{eq:trop_polyh211c} and marks its roots with red symbols.

Considering $\tilde{q}_{2,1,1}=0$, the characteristic polynomial and the tropical polynomial cast
\begin{align}
\BF_\lambda (t) =& \lambda^4 - (\tilde d_{21} t + \tilde d_{44}^2 t^2 + \tilde d_{34}\tilde d_{43}t^2)\lambda^2 \nonumber  \\& - (\tilde p_{2,1,1} t^2)\lambda ,\\
{\cal P}(t, \omega_{0}) =& \min \{ 4 \omega_{0}, 2 \omega_{0} +1, \omega_{0}+2\}.
\label{eq:trop_polyh211d}
\end{align}
The tropical roots are $\omega_{0}=1/2$ and $\omega_{0}=1$ with multiplicities $2$ and $1$, respectively. These roots are shown by red points in Fig.~\ref{fig:trop_poly_h211}~(d). Aside from a square-root dispersive doublet, the system hosts two more degeneracies, one with linear asymptotic dispersion and the other one is a non-dispersive eigenvalue at $\lambda=0$, which is not counted in the total multiplicities.

\paragraph{$H_{2,2}$ case. }

After examining the structure of the trivial perturbation in $H_{2,2}$, we obtain the general form of $H_{2,2}$ nontrivial perturbation as
\begin{align}
    H^{\rm tot}_{2,2}=H_{2,2}+\delta H_{2,2}=
    \left(
\begin{array}{cccc}
 0 & 1 & 0 & 0 \\
 d_{21} & 0 & d_{23} & d_{24} \\
 d_{31} & 0 & -d_{44} & 1 \\
 d_{41} & 0 & d_{43} & d_{44} \\
\end{array}
\right).
\end{align}
The associated characteristic polynomial reads
\begin{align}
  {\cal F}_{\lambda} &= \lambda ^4- \lambda ^2 \left(d_{21}+d_{43}+d_{44}^2\right)
  \nonumber \\
  &
  -\lambda  (d_{23} d_{31}+d_{24} d_{41})
    \nonumber \\
  &
  +d_{23} d_{31} d_{44}-d_{23} d_{41}  +d_{21} d_{43}
      \nonumber \\
  &
  -d_{24} d_{31} d_{43}-d_{24} d_{41} d_{44} +d_{21} d_{44}^2.
  \label{eq:charpolH22}
\end{align}
Parameterizing all variables $d_{ij}$ by $\tilde{d}_{ij} t$, with $t$ being a parameter along which the generic perturbation is performed, the characteristic polynomial can be rewritten as
\begin{align}
  {\cal F}_{\lambda} (t)&= \lambda ^4+ \lambda ^2 \left(-\tilde{d}_{21} t-\tilde{d}_{43} t-\tilde{d}_{44}^2 t^{2}\right)
  \nonumber \\
  &
  +\lambda  \tilde{r}_{2,2} t^{2}
  +\tilde{p}_{2,2} t^{2}
  +\tilde{q}_{2,2} t^{3}
  ,
\end{align}
where $\tilde{r}_{2,2} = (-\tilde{d}_{23} \tilde{d}_{31}-\tilde{d}_{24} \tilde{d}_{41})$, $\tilde{p}_{2,2}= \tilde{d}_{21} \tilde{d}_{43} -\tilde{d}_{23} \tilde{d}_{41}$, and $\tilde{q}_{2,2} = \tilde{d}_{23} \tilde{d}_{31} \tilde{d}_{44} +\tilde{d}_{21} \tilde{d}_{44}^2
  -\tilde{d}_{24} \tilde{d}_{31} \tilde{d}_{43}-\tilde{d}_{24} \tilde{d}_{41} \tilde{d}_{44}$.

\begin{figure}
      \centering
      \includegraphics[width=0.49\textwidth]{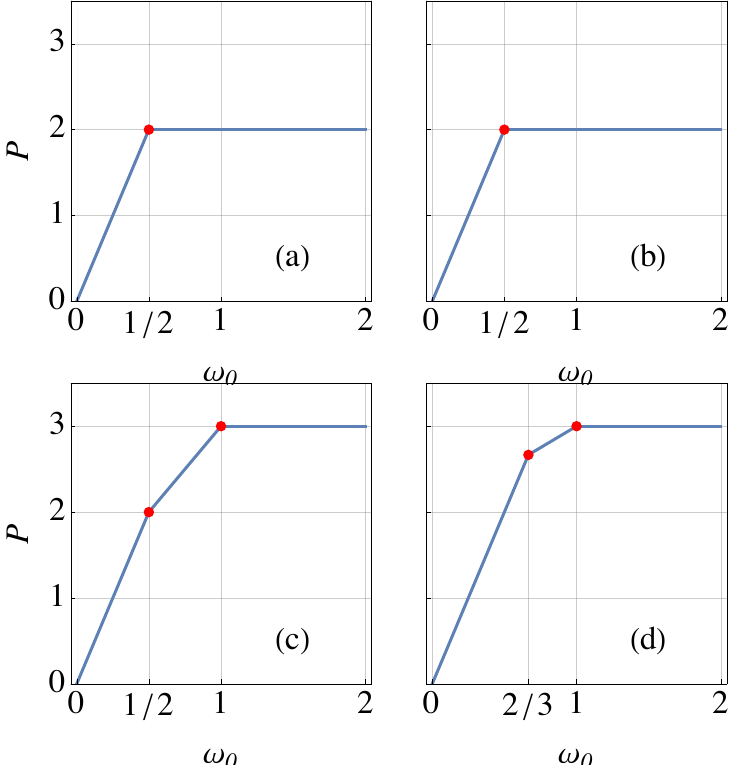}
      \caption{Possible tropical polynomials in $H_{2,2}$ with explicit forms given in Eq.~\eqref{eq:trop-poly_h22a}~(a), Eq.~\eqref{eq:trop-poly_h22b}~(b), Eq.~\eqref{eq:trop-poly_h22c}~(c),  and Eq.~\eqref{eq:trop-poly_h22d}~(d).
    Red points indicate the roots of the tropical polynomials.
    }
    \label{fig:trop_poly_h22}
  \end{figure}

  Then, the tropical polynomial associated with ${\cal F}$, plotted in Fig.~\ref{fig:trop_poly_h22}~(a), casts
  \begin{align}
      {\cal P} (t, \omega_{0}) & =
      \min\{4 \omega_{0},2 \omega_{0}+1,\omega_{0}+2,2 \}.
      \label{eq:trop-poly_h22a}
  \end{align}
  The roots of this polynomial are two $\omega_{0}=1/2$~(red points), where each of them has multiplicities of $2$. This is consistent with the Jordan form of $H_{2,2}$ with two EP2s.
  When first order terms in the coefficient of $\lambda$ vanish, i.e., $\tilde{d}_{21}=\tilde{d}_{43}=0$, the tropical polynomial, shown in Fig.~\ref{fig:trop_poly_h22}~(b), yields
        \begin{align}
      {\cal P} (t, \omega_{0}) & =
      \min\{4 \omega_{0},2 \omega_{0}+2,\omega_{0}+2,2 \}.
      \label{eq:trop-poly_h22b}
  \end{align}
  Then, we get a single tropical root $\omega_{0}=1/2$~(red points) with multiplicity $4$. This suggests that $H^{\rm tot}_{2,2}$ under this particular perturbation hosts four eigenvalues with square-root asymptotic dispersion. This is the same behavior we witnessed when $H^{\rm tot}_{2,2}$ is perturbed under generic perturbation in Fig.~\ref{fig:trop_poly_h22}~(a). Having $\tilde{p}_{2,2}=0$ brings the tropical polynomial into
    \begin{align}
      {\cal P} (t, \omega_{0}) & =
      \min\{4 \omega_{0},2 \omega_{0}+1,\omega_{0}+2,3 \}
     .
     \label{eq:trop-poly_h22c}
  \end{align}
  Figure~\ref{fig:trop_poly_h22}(c) displays this polynomial and marks its roots $\omega_{0}=\{ 1/2,1\}$, with multiplicities $2$ and $1$, respectively, by red points. This system allows the occurrence of two square-root dispersive eigenvalues and a single degeneracy with linear asymptotic dispersion. The fourth eigenvalue of this system $\lambda=0$ remains intact under this particular perturbation. When $\tilde{p}_{2,2}=0$ and $\tilde{d}_{21}=\tilde{d}_{43}=0$, the tropical polynomial reads
      \begin{align}
      {\cal P} (t, \omega_{0}) & =
      \min\{4 \omega_{0},2 \omega_{0}+2,\omega_{0}+2,3 \}
     ,
     \label{eq:trop-poly_h22d}
  \end{align}
with the nonzero tropical roots being $\omega_{0}=2/3$ and $\omega_{0}=1$ with multiplicities $3$ and $1$, respectively. We present Eq.~\eqref{eq:trop-poly_h22d} and its tropical roots~(red points) in Fig.~\ref{fig:trop_poly_h22}~(d). We emphasize that the exponent $2/3$ is anomalous: the Lidskii-Vishik-Ljusternik theory, which applies to generic perturbations, permits only exponents of the form $1/m$ with $m$ being the size of a Jordan block, so a splitting of the form $t^{2/3}$ cannot arise from any generic perturbation of $H_{2,2}$ and is a genuinely non-generic feature captured by our formulation.

\paragraph{$H_{3,1}$ case. }

\begin{figure}
      \centering
      \includegraphics[width=0.49\textwidth]{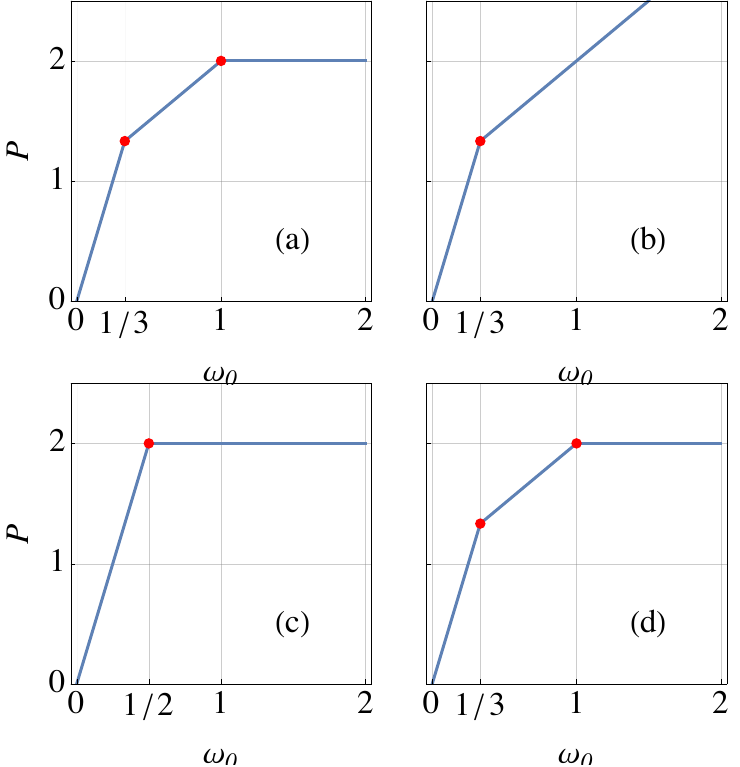}
      \caption{Possible tropical polynomials in $H_{3,1}$ with explicit forms given in Eq.~\eqref{eq:trop-poly_h31a}~(a), Eq.~\eqref{eq:trop-poly_h31b}~(b), Eq.~\eqref{eq:trop-poly_h31c}~(c),  and Eq.~\eqref{eq:trop-poly_h31d}~(d).
    Red points indicate the roots of the tropical polynomials. 
    }
    \label{fig:trop_poly_h31}
  \end{figure}

Following the procedure of generating trivial perturbations, we can deduce the general form of $H_{3,1}$ perturbed by nontrivial perturbation as
\begin{align}
    H^{\rm tot}_{3,1} =
      \left(
\begin{array}{cccc}
 0 & 1 & 0 & 0 \\
 0 & 0 & 1 & 0 \\
 d_{31} & d_{32} & -d_{44} & d_{34} \\
 d_{41} & d_{42} & 0 & d_{44} \\
\end{array}
\right),\label{eq:Htot31}
\end{align}
The associated characteristic polynomial reads
\begin{align}
 {\cal F}_{\lambda} &=\lambda ^4 +\left(d_{32} d_{44} -d_{34} d_{42} -d_{31}\right) \lambda 
 \nonumber \\
 &
 - \left(d_{44}^2 +d_{32}\right) \lambda ^2
 -d_{34} d_{41}+d_{31} d_{44}.\label{eq:chpol-H31}
\end{align}
Here, the only homogeneous polynomial in perturbation parameters is the coefficient of the free term, which is of order two. Performing the one-dimensional parametrization $d_{ij} =\tilde{d}_{ij} t$, we rewrite the characteristic polynomial in Eq.~\eqref{eq:chpol-H31} as
\begin{align}
\BF_\lambda(t) =& \lambda^4 - (\tilde{d}_{31}t - \tilde{d}_{32} \tilde{d}_{44} t^{2} +\tilde{d}_{34} \tilde{d}_{42} t^2)\lambda
\nonumber \\
&
- (\tilde{d}_{32}t+\tilde{d}_{44}^2 t^2) \lambda^2 
+ \tilde{q}_{3,1}t^2,
\end{align} 
where $\tilde{q}_{3,1} = \tilde{d}_{31} \tilde{d}_{44}-\tilde{d}_{34} \tilde{d}_{41}$.
The corresponding tropicalization, shown in Fig.~\ref{fig:trop_poly_h31}(a), is
\begin{align}
\BP(t,\omega_0) &= \min\{4\omega_0, 1+2\omega_0,1+\omega_0, 2\} .
\label{eq:trop-poly_h31a}
\end{align}
The two roots~(red points) are $\omega_{0}=1/3$ with multiplicity $3$ and a single $\omega_{0}=1$.
This is characteristic of a triplet of eigenvalues behaving like a cube root~(EP3) and one additional eigenvalue with linear asymptotic dispersion relation.

When $\tilde{q}_{3,1}=0$, the tropical polynomial casts
    \begin{align}
        {\cal P}(t, \omega_{0} ) = \min\{ 4 \omega_{0} , 2 \omega_{0}+1, \omega_{0}+1 \},
        \label{eq:trop-poly_h31b}
    \end{align}
    with its tropical roots being $\omega_{0}=1/2$ and $\omega_{0}=1$ associated with multiplicities $2$ and $1$, respectively. We present the tropical polynomial in Eq.~\eqref{eq:trop-poly_h31b} and its roots~(red points) in Fig.~\ref{fig:trop_poly_h31}(b). The spectrum of this perturbed Jordan block exhibits two degeneracies with square-root dispersion, an eigenvalue with linear asymptotic dispersion and an unperturbed eigenvalue at $\lambda=0$.
    Having $\tilde{d}_{31}=0$, the tropical polynomial, plotted in Fig.~\ref{fig:trop_poly_h31}(c), becomes
     \begin{align}
        {\cal P}(t, \omega_{0} ) = \min\{ 4 \omega_{0} , 2 \omega_{0}+1, \omega_{0}+2,2 \}.
        \label{eq:trop-poly_h31c}
    \end{align}
    This polynomial has two $\omega_{0}=1/2$ tropical roots, each with multiplicity $2$. Red points in Fig.~\ref{fig:trop_poly_h31}(c) indicate these tropical roots. Hence, with $\tilde{d}_{31}=0$, a doublet of degeneracies with square-root asymptotic dispersion find room to emerge in the spectrum of $H^{\rm tot}_{3,1}$.

     When $\tilde{d}_{32}=0$, the tropical polynomial reads
     \begin{align}
        {\cal P}(t, \omega_{0} ) = \min\{ 4 \omega_{0} , 2 \omega_{0}+2, \omega_{0}+1,2 \}.
        \label{eq:trop-poly_h31d}
    \end{align}
   The nonzero tropical roots are then $\omega_{0}=1$ and $\omega_{0}=1/3$ with multiplicities of $1$ and $3$, respectively. Figure~\ref{fig:trop_poly_h31}(d) exhibits the tropical polynomial in Eq.~\eqref{eq:trop-poly_h31d} and its tropical roots~(red points). In this system, we have three eigenvalues with cubic root asymptotic dispersions~(EP3) and a single eigenvalue with linear leading order.

\paragraph{$H_{4}$ case. }

\begin{figure}
      \centering
      \includegraphics[width=0.499\textwidth]{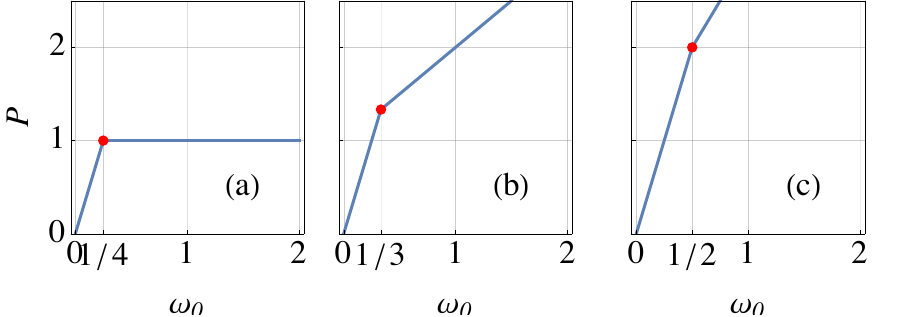}
      \caption{Possible tropical polynomials in $H_{4}$ with explicit forms given in Eq.~\eqref{eq:trop-poly_h4a}~(a), Eq.~\eqref{eq:trop-poly_h4b}~(b), and Eq.~\eqref{eq:trop-poly_h4c}~(c).
    Red points indicate the roots of the tropical polynomials.
    }
    \label{fig:trop_poly_h4}
  \end{figure}

The general nontrivial perturbation for $H_{4}$ Jordan form reads~\cite{Arnold1971, Sayyad2022}
\begin{align}
     H^{\rm tot}_{4} =
      \left(
\begin{array}{cccc}
 0 & 1 & 0 & 0 \\
 0 & 0 & 1 & 0 \\
 0 & 0 & 0 & 1 \\
 d_{41} & d_{42} & d_{43} & 0 \\
\end{array}
\right). \label{eq:Htot4}
\end{align}
The associated characteristic polynomial for $H^{\rm tot}_{4}$ is
\begin{align}
  {\cal F}_{\lambda} &= \lambda ^4 -d_{43} \lambda ^2-d_{42} \lambda -d_{41} 
.\label{eq:chpol4}
\end{align}
Defining the variable $t$ along fixed one-dimensional complex linear subspace, we rewrite the characteristic polynomial~\eqref{eq:chpol4} as
\begin{equation}
\BF_\lambda (t) = \lambda^4 - (\tilde{d}_{43} t) \lambda^2 - (\tilde{d}_{42} t)\lambda - \tilde{d}_{41} t.
\end{equation}
For the general direction, when all the coefficients are nonzero, the corresponding tropicalization reads
\begin{align}
\BP(t,\omega_0) &= \min\{4\omega_0, 1+2\omega_0, 1+\omega_0,1\}.
\label{eq:trop-poly_h4a}
\end{align}
Here, there is a single non-zero tropical root $1/4$ with multiplicity $4$, typical for EP4s; see Fig.~\ref{fig:trop_poly_h4}(a).

Note that the tropical polynomial does not care whether $\tilde{d}_{43}$, $\tilde{d}_{42}$ and $\tilde d_{42}$ vanish as long as $\tilde{d}_{41}$ is non-zero. As for $H^{\rm tot}_{4}$, we generally only care about the non-zero coefficients at the smallest power of $\lambda$.
When $\tilde{d}_{42}\neq0$ while $\tilde{d}_{41}=0$, then
\begin{equation}
\BP(t,\omega_0) = \min\{4\omega_0,1+\omega_0\}.
\label{eq:trop-poly_h4b}
\end{equation}
There is a non-zero root $1/3$ with multiplicity $3$; see Fig.~\ref{fig:trop_poly_h4}(b). Since there are $4$ eigenvalues, the fourth one has the leading power $0$. Hence, this system hosts three cubic-root-dispersive degeneracies and a flat unperturbed eigenvalue $\lambda=0$.

Finally, if only $\tilde{d}_{43}$ is non-zero,
then
\begin{equation}
\BP(t,\omega_0) = \min\{4\omega_0,1+2\omega_0\}.
\label{eq:trop-poly_h4c}
\end{equation}
There is a single non-zero root $1/2$ of multiplicity $2$; see Fig.~\ref{fig:trop_poly_h4}(c). Hence, the system has two eigenvalues with square-root asymptotic dispersion, and the other two eigenvalues~($\lambda=0$) should have leading power $0$.

We note that the current tropical polynomial reproduces the results of the Lidskii-Vishik-Ljusternik theorem for general perturbation directions in a simple and yet rigorous way.

We summarize emerging tropical roots and their multiplicities for generic perturbations of Jordan forms of
$4 \times 4$ matrices in Table~\ref{tab:rank4}~\footnote{Some of our presented degeneracies under generic perturbations are also reported in Ref.~\cite{Bid2025} where leading perturbative behavior of eigenvalues are explored using the response theory.}.

\section{Examples}\label{sec:examples}

In this section, we present various examples that were extensively explored in various research directions to demonstrate the wide applicability range of our findings.

\subsection{Exceptional points and enhanced sensing}
The research direction of EP-based sensing has received immense attention in recent years~\cite{Wiersig2014, Liu2016,  Hodaei2017, Djorwe2019, Zhang2019, Xiao2019, Budich2020, Wiersig2020, Wu2021, Wong2023, Shi2025}. These studies suggested that the presence of higher-order EPs gives rise to the development of advanced sensors. In the following, we present multiple examples of EP-based sensing systems.

\paragraph{Cavity-mediated atomic sensors. }
An example of such systems is cavity-mediated atomic sensors, where dynamical phase transitions between parametric amplification of dynamical behavior and periodic oscillation in the time evolution of the system give rise to the emergence of higher-order EPs. 

The magnon-cavity systems are governed by a dynamical matrix that exhibits symmetries such as particle-hole symmetry and pseudo-Hermiticity~\cite{Shi2025}. The spontaneous breaking of particle-hole symmetry gives rise to the emergence of EPs in these systems. The generic structure of dynamical matrix describing atom-cavity systems reads
\begin{align}
   d_{m,n} = \begin{pmatrix}
        \delta_{1} &  \ldots & 0 &0& \ldots& 0 & g_{1} \\
        \vdots & \ddots & \vdots & \vdots&\ddots& \vdots & \vdots \\
        0 &  \ldots & \delta_{m} & 0& \ldots& 0 & g_{m} \\
        0 &  \ldots & 0 & -\delta_{m+1}& \ldots& 0 & -\kappa_{1} \\
         \vdots & \ddots & \vdots & \vdots&\ddots& \vdots & \vdots \\
         0 &  \ldots & 0 & 0& \ldots& -\delta_{n+m-1} & -\kappa_{n-1} \\
         -g_{1} &  \ldots & -g_{m} &-\kappa_{1}& \ldots& -\kappa_{n-1}  & 0 \\
    \end{pmatrix},
\end{align}
where the size of $d_{m,n}$ is $n+m$. Here, $\delta_{i}$ accounts for two-photon detuning and external perturbations on a particular magnon mode $i$~\footnote{Magnons are quasiparticles describing the spin-wave excitations in atomic systems.}. Two coupling constants $g_{i}$ and $\kappa_{i}$ denote two different types of magnon-cavity interactions; see further details in Ref.~\cite{Shi2025}. The matrix $d_{m,n}$ satisfies the pseudo-Hermiticity condition, $d_{m,n}= \xi d_{m,n}^{\dagger} \xi^{-1}$ with $\xi= {\rm diag}[-1, \ldots, -1, 1, \ldots, 1]$ where the number of $-1(1)$ eigenvalues are $m(n)$.

To demonstrate how our proposed approach facilitates characterizing EPs, let us consider two cases $(m,n)=(1,2)$ and $(m,n)=(2,2)$. We set the matrix elements in the dynamical matrix $d_{1,2}$ such that
\begin{align}
    d_{1,2}= \left(
\begin{array}{ccc}
 -\gamma  & 0 & 1 \\
 0 & \gamma  & -1 \\
 -1 & -1 & 0 \\
\end{array}
\right),
\end{align}
with $\gamma$ being the perturbation parameter.
At $\gamma =0$, this dynamical matrix hosts an EP3.
The characteristic polynomial for $d_{1,2}$ reads
\begin{equation}
     {\cal F}_{\lambda}(\gamma) =  \lambda ^3 - \gamma ^2 \lambda +2\gamma.
\end{equation}
Tropicalizing the above equation with respect to $\gamma$ gives
\begin{align}
    {\cal P}(\gamma, \omega_{0}) = \min \{ 3 \omega_{0}, \omega_{0} +2 ,1 \},
\end{align}
with a single tropical root $\omega_{0}=1/3$ with multiplicity $3$. Looking at the spectrum of this system with eigenvalues 
\begin{align}
    \lambda_{1} &= -e^{i \pi/3} f_{+} - e^{-i\pi/3} f_{-},\\
    \lambda_{2} &=  -e^{-i \pi/3} f_{+} - e^{i\pi/3} f_{-},\\
    \lambda_{3} &= f_{+} +f_{-},
\end{align}
where $f_{\pm}= (-\gamma \pm 2 \gamma \sqrt{\gamma^{4} - 27} )^{1/3}$, unsurprisingly confirms the presence of eigenvalues with asymptotic dispersion $\gamma^{1/3}$ resulting from the splitting of an EP3.

Let us now explore $d_{2,2}$ with two sets of parameters. First, consider 
\begin{align}
    d_{2,2} =
    \left(
\begin{array}{cccc}
 -\gamma  & 0 & 0 & 1 \\
 0 & \frac{\gamma}{2} & 0 & 0 \\
 0 & 0 & \gamma  & -1 \\
 -1 & 0 & -1 & 0 \\
\end{array}
\right),
\end{align}
where $\frac{\gamma}{2}$ is clearly one of its eigenvalues with linear asymptotic behavior in $\gamma$.
For $\gamma=0$, $d_{2,2}$ hosts an EP(3,1).
The characteristic equation for this matrix reads
\begin{equation}
     {\cal F}_{\lambda}(\gamma) =  \lambda ^4-\frac{\lambda^3 \gamma }{2} -\lambda^2 \gamma^2+\frac{1}{2} \lambda  \left(\gamma ^3-4 \gamma \right)+\gamma^2,
     \label{eq:Fd24_eig31}
\end{equation}
with its tropical polynomial being
\begin{align}
    {\cal P}(\gamma, \omega_{0}) = \min \{4 \omega_{0}, 3 \omega_{0}+1, 2\omega_{0} +2 , \omega_{0}+1,2 \}.
\end{align}
The tropical roots are $\omega_{0}=1$ and $\omega_{0}=1/3$ with algebraic multiplicity $1$
 and $3$, respectively. We have already verified the $\omega_{0}=1$ associated with the eigenvalue $\lambda_{1}=\frac{\gamma}{2}$. Removing the contribution of this eigenvalue from Eq.~\eqref{eq:Fd24_eig31}, we obtain $\lambda ^3 - \gamma ^2 \lambda -2\gamma$. The asymptotic behaviors of eigenvalues satisfying this equation read~\cite{Shi2025}
 \begin{align}
     \lambda_{2} &= e^{-2\pi i /3} (2 \gamma)^{1/3} ,\\
     \lambda_{3} &= e^{2\pi i /3} (2 \gamma)^{1/3} ,\\
     \lambda_{4} &=  (2 \gamma)^{1/3} ,
 \end{align}
which is reminiscent of EP3 splitting with tropical root $\omega_{0}=1/3$.

\begin{figure}
    \centering
    \includegraphics[width=0.99\linewidth]{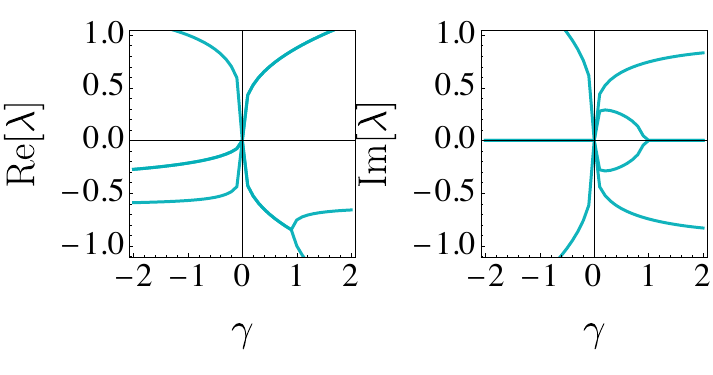}
    \caption{Eigenvalues of $d_{2,2}$ given in Eq.~\eqref{eq:d24_ep4} exhibiting EP4s with $\gamma^{1/4}$ asymptotic behavior close to $\gamma=0$. }
    \label{fig:d24_ep4}
\end{figure}

The second set of parameters of our interests to form $d_{2,2}$ results in
\begin{align}
    d_{2,2}= \left(
\begin{array}{cccc}
 \sqrt{\gamma } & 0 & 0 & \frac{\sqrt{\sqrt{\gamma }+1}}{\sqrt{2}} \\
 0 & -\sqrt{\gamma } & 0 & \frac{\sqrt{1-\sqrt{\gamma }}}{\sqrt{2}} \\
 0 & 0 & 0 & -1 \\
 -\frac{\sqrt{\sqrt{\gamma }+1}}{\sqrt{2}} & -\frac{\sqrt{1-\sqrt{\gamma }}}{\sqrt{2}} & -1 & 0 \\
\end{array}
\right),\label{eq:d24_ep4}
\end{align}
with the characteristic polynomial being
\begin{align}
    {\cal F}_{\lambda}(\gamma) = \lambda^4 -\gamma  \lambda ^2+\gamma  \lambda +\gamma .
\end{align}
Tropicalizing this equation then reads
\begin{align}
    {\cal P}(\gamma, \omega_{0})= \min\{
 4\omega , 2 \omega_{0}+1 ,\omega_{0}+1 , 1 \},
\end{align}
where the only tropical root is $\omega_{0}=1/4$ with multiplicity $4$ typical for EP4 splitting; see also the dispersion relation near $\gamma=0$ in Fig.~\ref{fig:d24_ep4}.

\paragraph{Electronic sensing circuit. }
Electronic circuits are another set of systems that perform as higher-order EP-based sensors. A nice experimental example of such systems hosts EP6 or EP4 in the presence of certain perturbations~\cite{Xiao2019, Xiao2023}. 

The circuit Laplacian~($\cal L$) of a system with two resonators, each with a capacitor and inductor, coupled resistors, and a capacitor reads~\footnote{See Ref.~\cite{Xiao2019} for further details on the circuit design.}
\begin{align}
    {\cal L}= 
    \left(
\begin{array}{cccccc}
 \epsilon & 0 & 0 & i & 0 & 0 \\
 0 & 0 & 0 & 0 & i & 0 \\
 0 & 0 & 0 & 0 & 0 & i \\
 -i & i & 0 & -i \gamma  & 0 & 0 \\
 i \mu  & -2 i \mu  & i \mu  & 0 & 0 & 0 \\
 0 & i & -i & 0 & 0 & i \gamma  \\
\end{array}
\right),\label{eq:Lcirc1}
\end{align}
where the coupling between two resonators is set by $\mu$, $\gamma$ denotes the loss and gain rate in each resonator, and $\epsilon$ is the perturbation parameter. Having the input current $I$ and node voltage $V$, the system satisfies $I= {\cal L} V$.
For $\epsilon=0$, the matrix hosts an EP6.
The corresponding characteristic polynomial reads
\begin{align}
   {\cal F}_{\lambda}=& 
   \lambda ^6-\lambda ^5 \epsilon -\lambda ^3 \epsilon +\lambda ^2 \left(\frac{i \epsilon }{2}+\frac{1}{2} i \sqrt{5} \epsilon \right)
   \nonumber \\
   &
   +\lambda  \left(\frac{\sqrt{5} \epsilon }{4}+\frac{3 \epsilon }{4}\right)-\frac{i \epsilon }{2}.
\end{align}
At the critical values $\gamma_{\rm EP}=\frac{1}{2} \left(\sqrt{5}+1\right)$ and $ \mu_{\rm EP}=\frac{1}{4} \left(\sqrt{5}-1\right)$, the tropical polynomial casts
\begin{multline}
    {\cal P}(\epsilon, \omega_{0})= \\\min\{6 \omega_{0},5 \omega_{0}+1,3 \omega_{0}+1,2 \omega_{0}+1,\omega_{0}+1,1\}.
\end{multline}
The single tropical root is $\omega_{0}=1/6$ with algebraic multiplicity $6$, which agrees with the experimental report in Ref.~\cite{Xiao2019}. 

Setting $\epsilon=0$ in Eq.~\eqref{eq:Lcirc1}, $\mu=\mu_{\rm EP}$, and introducing perturbation of $\gamma$ such that $\gamma = \eta + \gamma_{\rm EP}$, with $\eta$ being small, we obtain the following characteristic polynomial for the system
\begin{align}
      {\cal F}_{\lambda}=& \lambda ^6+\lambda ^4 \eta  \left(\eta +\sqrt{5}+1\right)
      \nonumber \\
      &
      +\frac{1}{2} \lambda ^2 \eta  \left(-\sqrt{5} \eta +\eta -4\right).
\end{align}
Tropicalizing this equation reads
\begin{align}
    {\cal P}(\eta, \omega_{0})=
    \min\{6 \omega_{0},4 \omega_{0}+1,2 \omega_{0}+1\},
\end{align}
with its root being $\omega_{0}=1/4$ and multiplicity $4$. Hence, the eigenvalues exhibit asymptotic dispersion reminiscent of an EP4 instead of an EP6.

\subsection{Exceptional points in nonreciprocal systems}

One of the prominent examples of non-Hermitian systems is the nonreciprocal system with imbalanced gain/loss, or asymmetric hopping amplitudes. The nonreciprocity in these systems may give rise to the appearance of skin effects, non-Hermitian phenomena with no Hermitian counterparts~\cite{Zhang2022review}. To implement our framework on nonreciprocal systems, let us consider a paradigmatic bipartite Hatano-Nelson model. Here, the quadratic Hamiltonian on a one-dimensional system with size $L$ reads
\begin{align}
    {\cal H}=
    \begin{pmatrix}
 0 & t_{1}-\gamma_{1} & 0&\ldots  & 0 & \eta_{1} \\
t_{1}+ \gamma_{1} & 0 & t_{2}-\gamma_{2} &  \ldots & 0  & 0 \\
 0 & t_{2} + \gamma_{2}  & 0 &  \ldots & 0 &0 \\
 \vdots & \vdots & \vdots & \ddots& \vdots & \vdots  \\
 0 & 0 & 0 & 0 & 0 & c_{1}  \\
 \eta_{2} & 0 & 0 & 0 & c_{2}& 0 \\
    \end{pmatrix}_{L \times L},
    \label{eq:Hnonrecip}
\end{align}
where $c_{1/2} = t_{1} -\pm  \gamma_{1}$ when $L$ is odd and $c_{1/2} = t_{2} -\pm  \gamma_{2}$ for even values of $L$. 

To describe systems with open boundary conditions, we set $\eta_{1}=\eta_{2}=0$.
A particular parameter regime of our interest is when $t_{2}=\gamma_{2}=1$ and $t_{1}= \gamma_{1}+\epsilon$, with $\epsilon$ being a perturbative parameter. In this case, the characteristic polynomial casts
\begin{align}
    {\cal F}_{\lambda} = \lambda^{{\rm mod}[L, 2]} \left( \lambda^{2} - \epsilon(2 \gamma_{1} + \epsilon) \right)^{\lfloor L/2 \rfloor},
\end{align}
where $\lfloor x \rfloor$ is the floor of $x$. It is evident that the system hosts a flat $\lambda=0$ eigenvalue when $L$ is odd. Tropicalizing each of the remaining polynomial factors, namely, $ \lambda^{2} - \epsilon(2 \gamma_{1} + \epsilon)$, for any value of $\gamma_{1}$ with respect to $\epsilon$ results in 
\begin{align}
    {\cal P}(\epsilon, \omega_{0}) = \min\{2 \omega_{0}, 1\},
\end{align}
with a single root $\omega_{0}=1/2$ with algebraic multiplicity $2$ associated with EP2s. Hence, the system hosts $2{\lfloor L/2 \rfloor}$ EP2s when $L$ is even, and it exhibits an extra dispersionless eigenvalue $\lambda=0$ for odd values of $L$.

Another interesting parameter regime is when $\eta_{1}=0$ and $\eta_{2}= \epsilon$ which sets perturbative unidirectional hopping amplitudes between the first and the last site of the system. When $t_{1}=-\gamma_{1}$ and $t_{2}=-\gamma_{2}$, the lower diagonal elements in Eq.~\eqref{eq:Hnonrecip} vanishes and the systems acquires a structure similar to Jordan normal forms, with an extra $\epsilon$ element. The generic form of characteristic polynomial for this system then reads
\begin{align}
    {\cal F}_{\lambda} = \lambda^{L} - 2^{L-1} \, t_{1}^{\lfloor L/2 \rfloor} \, t_{2} ^{\lfloor (L-1)/2 \rfloor} \, \epsilon.
\end{align}
For any nonzero values of $t_{1}$ and $t_{2}$, tropicalizing the above relation with respect to $\epsilon$ gives
\begin{align}
    {\cal P}(\epsilon, \omega_{0}) = \min\{L \omega_{0}, 1\},
\end{align}
where the tropical root is $\omega_{0}=1/L$ with algebraic multiplicity $L$, the well-known signature of splitting EP$L$s.

\subsection{Exceptional points in non-Hermitian lattice models}
The other prominent platforms where non-Hermitian degeneracies play a significant role in giving rise to various exotic phenomena~\cite{Wangwang2022, Sayyad2022c, Sticlet2023, Legal2023, turkeshi2023entanglement, MaKou2024, Sayyad2024, Du2025, Chaturvedi2025} are condensed matter systems, particularly the spectrum of non-Hermitian lattice models. To demonstrate the idea of how our approach facilitates identifying asymptotic spectral behavior close to non-Hermitian degeneracies, we consider a non-Hermitian Lieb's model~\footnote{Other varieties of non-Hermitian Lieb's model are explored in \cite{Mandal2021, Bid2025}.}. 

\begin{figure}[t!]
    \centering
    \includegraphics[width=0.85\linewidth]{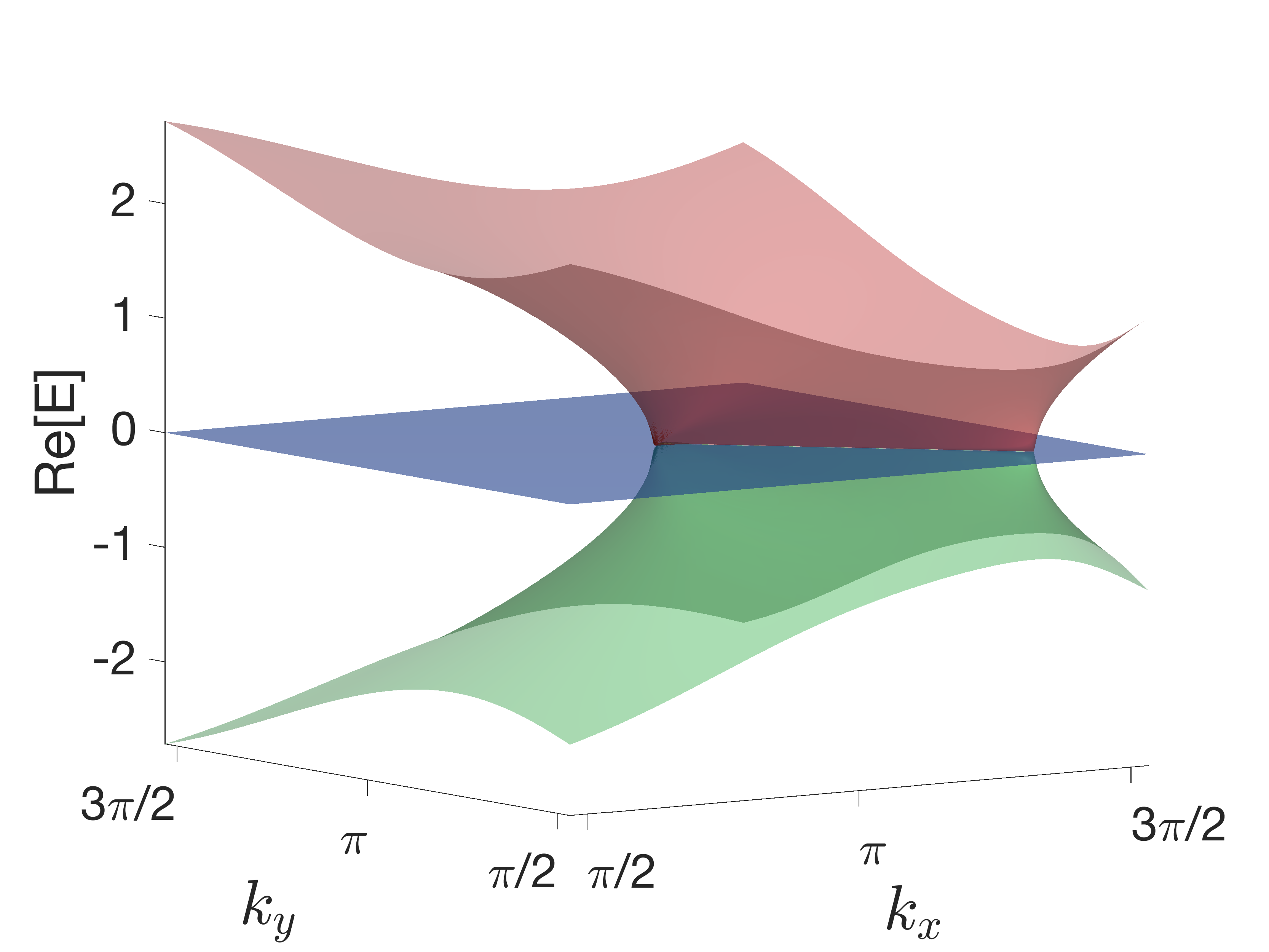}
    \includegraphics[width=0.85\linewidth]{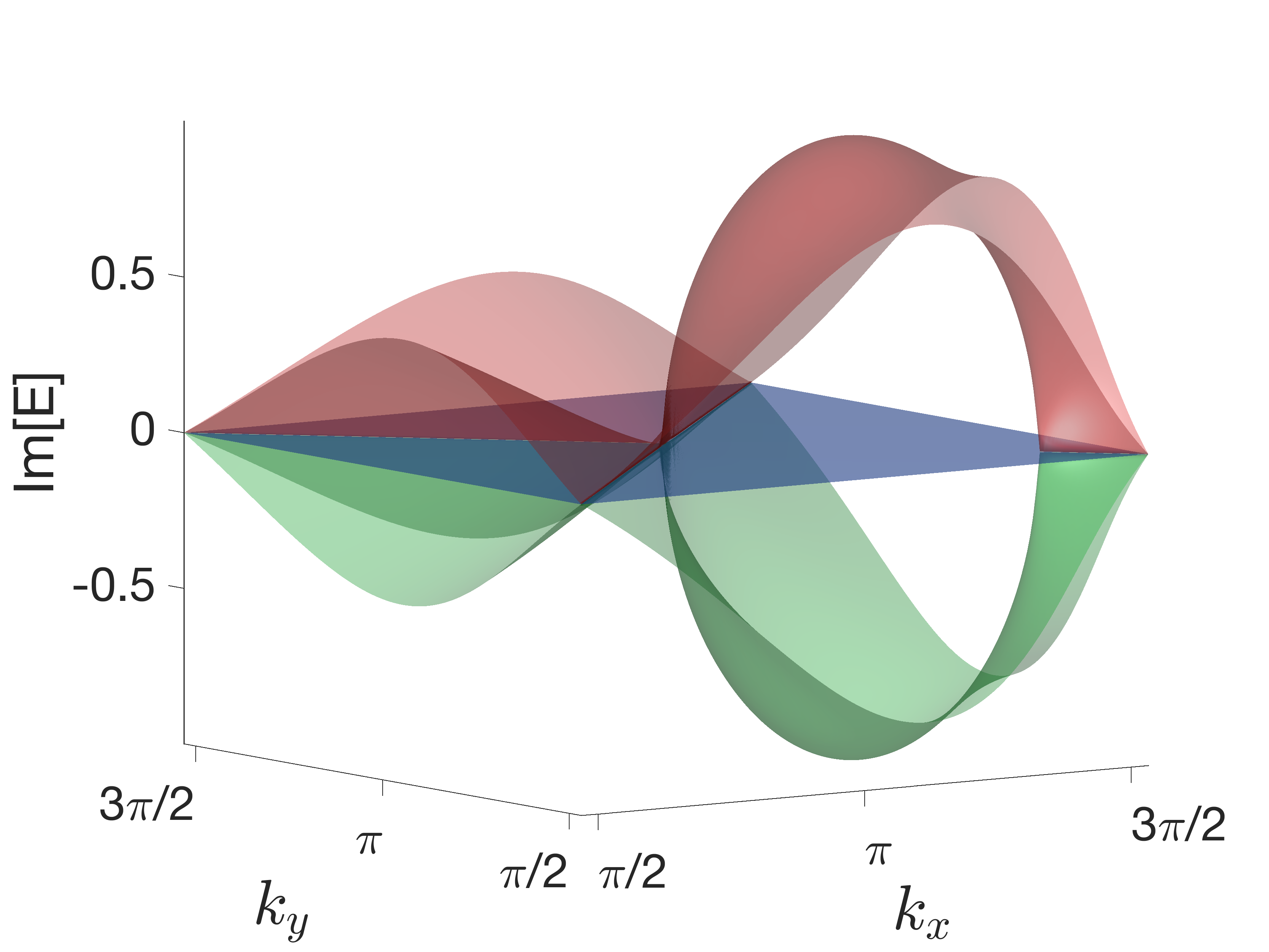}
    \caption{Real~(top panel) and imaginary~(bottom) parts of the spectrum of the NH Lieb's model in Eq.~\eqref{eq:Hlieb} at $\epsilon=1.5$.}
    \label{fig:spectrum_nhlieb}
\end{figure}

The three-band model Hamiltonian on the Lieb lattice reads
\begin{align}
   {\cal H} =  \left(
\begin{array}{ccc}
 0 & 1+e^{i k_y} & 0 \\
 1+e^{-i k_y}+i \epsilon  & 0 & 1+e^{-i k_x}-i \epsilon  \\
 0 & 1+e^{i k_x} & 0 \\
\end{array}
\right),
\label{eq:Hlieb}
\end{align}
where $k_{x}$ and $k_{y}$ are, respectively, lattice momentum along the $x$ and $y$ directions and $\epsilon$ denotes the non-Hermitian parameter. The spectrum of this Hamiltonian has one flat band with energy $E=0$ and two dispersive bands given by
\begin{align}
  E= \pm    \sqrt{\epsilon  \sin (k_x) -\epsilon  \sin (k_y) + A},
  \label{eq:Elieb}
\end{align}
with $A= (2-i \epsilon ) \cos (k_x)+(2+i \epsilon ) \cos (k_y)+4$; an example of the spectrum at $\epsilon=1.5$ is shown in Fig.~\ref{fig:spectrum_nhlieb}. Two non-Hermitian degeneracies occur at $k_y=-k_x= 2 \arccot(\epsilon/2)$ and $k_x=k_y=\pi$.

\begin{figure}[t!]
    \centering
    \includegraphics[width=0.95\linewidth]{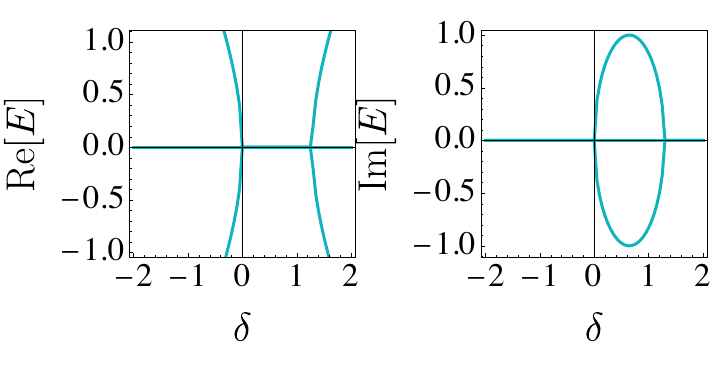}
    \caption{The real and imaginary part of the NH Lieb's model close to $k_y=-k_x=2 \delta +2 \cot ^{-1}\left(\frac{\epsilon }{2}\right)$ at $\epsilon=1.5$. Note that there is a flat band at $E=0$.}
    \label{fig:nh_lieb_arccot}
\end{figure}

To understand the asymptotic dispersion in the vicinity of the first degeneracy, we introduce a perturbation parameter $\delta$ along the $k_x=-k_y$ line close to $k_y=-k_x= 2 \cot ^{-1}(\epsilon/2)$. The associated characteristic polynomial reads
\begin{align}
    \lambda ^3+ \lambda  (4 \cos (\delta )-4)=0,
\end{align}
where after approximating $ \cos (\delta )= 1+ \delta^/2$, its tropicalization is
\begin{align}
    P_{\delta}(\omega_{0}) = \min\{3 \omega_{0},\omega_{0}+2\}.
\end{align}
The root of the above tropical polynomial is $\omega_{0}=\frac{1}{2}$ with algebraic multiplicity $2$; See the spectrum in Fig.~\ref{fig:nh_lieb_arccot}. To identify the order of our degeneracy, we notice that the second power of Hamiltonian, specifically $({\cal H}-\lambda I)^{2}$ with $I$ being an identity matrix and $\lambda=0$, at $k_y = -k_x = 2 \cot^{-1}(\epsilon/2)$ is not a null matrix while $({\cal H}-\lambda I)^{3}$ form a null matrix. This suggests that an EP3 resides at $k_y = -k_x = 2 \cot^{-1}(\epsilon/2)$, and the perturbation we have considered is non-generic. 

\begin{figure}[t!]
    \centering
    \includegraphics[width=0.95\linewidth]{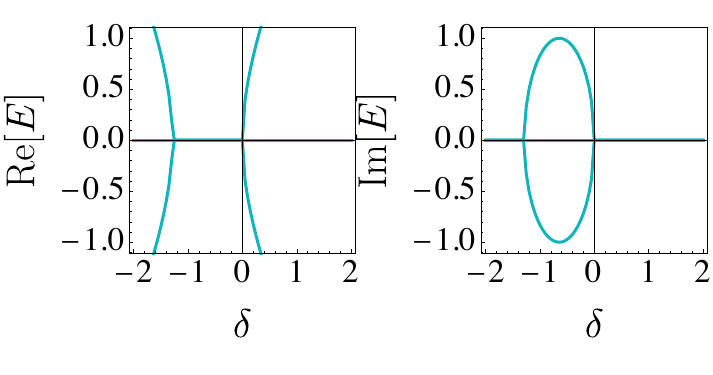}
    \caption{The real and imaginary part of the NH Lieb's model along $k_x=\pi+\delta$ and $k_x=\pi-\delta$ at $\epsilon=1.5$. Note that there is a flat band at $E=0$.}
    \label{fig:nh_lieb_piquadratic}
\end{figure}

Perturbing the second degeneracy at $k_x=k_y=\pi$ along $k_x=-k_y$ with parameter $\delta$ results in the following characteristic polynomial.
\begin{align}
   \lambda ^3+\lambda  (2 \epsilon  \sin (\delta )+4 \cos (\delta )-4)=0.
\end{align}
Approximating $\sin(\delta)$ and $\cos(\delta)$ by their smallest series expansions and tropicalizing the approximated characteristic polynomial, then reads
\begin{align}
    P_{\delta}(\omega_{0})= \min \{3 \omega_{0},\omega_{0}+1\},
\end{align}
with its root being $\omega_{0}=1/2$ with multiplicity $2$. Hence, the system exhibits another EP2 intersecting with a nondispersive band at $E=0$; see the low-energy spectrum of the model in Fig.~\ref{fig:nh_lieb_piquadratic}.

\begin{figure}[t!]
    \centering
    \includegraphics[width=0.95\linewidth]{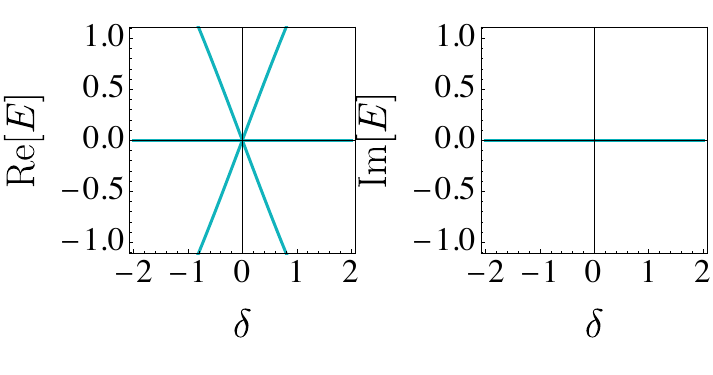}
    \caption{The real and imaginary part of the NH Lieb's model along $k_x=k_y=\pi+\delta$ at $\epsilon=1.5$. Note that there is a flat band at $E=0$.}
    \label{fig:lieb_pilinear}
\end{figure}

We note that the spectrum in Eq.~\eqref{eq:Elieb} is purely real along $k_{x}=k_y$. Perturbing the degeneracy at $k_x=k_y=\pi$ along $k_x=k_y$ does not affect the realness of the spectrum; see Fig~\eqref{fig:lieb_pilinear}.
The associated characteristic polynomial of the model in the limit of small $\delta$ reads
\begin{align}
   \lambda ^3- 2 \delta ^2 \lambda  =0,
\end{align}
corresponding to the tropical polynomial
\begin{align}
    P_\delta(\omega_{0})=\min\{ 3 \omega_{0}, \omega_{0}+2\},
\end{align}
with its root being $\omega_{0}=1$ with multiplicity 2.
Hence, the system exhibits two linearly dispersive bands and a flat band in the vicinity of $k_{x}=k_{y}=\pi$ and along $k_{x}=k_{y}$. At the same time, $(\BH-\lambda I)^2$ already forms a zero matrix, which indicates the presence of EP(2,1) at $k_x=k_y=\pi$. Again, the perturbation we considered was non-generic.

\subsection{Exceptional points of the Liouvillian superoperator}

Typically, non-Hermitian physics is considered as an approximation of an open system. The dynamics of these open systems is described under general assumptions by the Lindblad master equation
\begin{multline}
    \frac{d}{dt}\hat \rho = \BL\qty[\hat \rho] = \\-i\qty[\hat H,\hat\rho]+\sum_{k}\Gamma_k\qty[\hat L_k \hat \rho \hat L_k^\dagger - \frac{1}{2}\{\hat L^\dagger_k\hat L_k,\hat\rho\}]\label{lindblad}.
\end{multline}
Here, $\hat \rho$ and $\hat H$ are the density matrix and the Hermitian Hamiltonian of a system, respectively, while $\hat L_k$ is a set of jump operators describing the influence of a bath. Any operator on the Hilbert space $\BH$ can be identified with a vector in the tensor product $\BH\otimes\BH^*$,
which allows us to vectorize the density matrix
\begin{equation}
 \hat \rho = \sum_{m,n} c_{m,n}\ket{m}\bra{n}\longrightarrow \bar\rho = \sum_{m,n} c_{m,n} \ket{m}\otimes\ket{n^*}.
\end{equation}
In this vectorized representation, Liouvillian $\BL\qty[\hat\rho]$ simply becomes a matrix
\begin{align}
    \frac{d}{dt}\bar\rho &=\BL\bar \rho,\\
    \BL &= (-i\hat H_\mathrm{nh})\otimes\mathbb1 + \mathbb1\otimes(i\hat H_\mathrm{nh}^*) + \sum_k \Gamma_k \hat L_k\otimes \hat L_k^*,
\end{align}
where 
\begin{equation}
    \hat H_\mathrm{nh} = \hat H - \frac{1}{2}\sum_k \Gamma_k \hat L_k^\dagger\hat L_k.
\end{equation}
Usually, non-Hermitian physics is associated with $\hat H_\mathrm{nh}$. However, the Liouvillian superoperator $\BL$ is also a non-Hermitian matrix and can host EPs.

Even if the quantum jump terms $\Gamma_k \hat L_k\otimes \hat L_k^*$ are neglected, transition from the level of the non-Hermitian Hamiltonian to the non-Hermitian Liouvillian
\begin{equation}
    \BL^\prime = (-i\hat H_\mathrm{nh})\otimes\mathbb1 + \mathbb1\otimes(i\hat H_\mathrm{nh}^*)
\end{equation}
is rather non-trivial. For example, if the non-Hermitian Hamiltonian is tuned to an EP of the maximal order $N$ with eigenvalue $\varepsilon$ as
\begin{equation}
    \hat H_\mathrm{nh}\sim J_N(\varepsilon),
\end{equation}
then the non-Hermitian Liouvillian naturally exhibits a multiblock EP~\cite{shiralieva2025, trampus1966}:
\begin{equation}
    \BL^\prime \sim J_{2N-1}(\lambda_\mathrm{EP})\oplus J_{2N-3}(\lambda_\mathrm{EP})\oplus\dots\oplus J_1(\lambda_\mathrm{EP}),
\end{equation}
where $\lambda_\mathrm{EP} = i(\varepsilon^*-\varepsilon)=2\im \varepsilon$. This proves to be important even when the quantum jump terms are taken into account.

As a concrete example, let us consider a four-level system with the Hamiltonian
\begin{equation}
    \hat H = \sum_{l=1}^4 \varepsilon_l \ket{l}\bra{l} + \sum_{l<k}^{4} t_{lk}(\ket{l}\bra{k}+\ket{k}\bra{l}),
\end{equation}
subject to dissipation from the upper levels to the lower ones:
\begin{align}
    \BL&=\qty(-i\hat H)\otimes\mathbb1+\mathbb1\otimes\qty(i\hat H) + \sum_{l<k}\Gamma_{lk} D\qty[\ket{l}\bra{k}],\\
    D\qty[\hat L] &= \hat L\otimes\hat L^* - \frac{\hat L^\dagger\hat L\otimes\mathbb1 + \mathbb1\otimes\hat L^T\hat L^*}{2}.
\end{align}
Under the assumption that there is no coherent coupling to the ground state $t_{1k}\equiv 0$, we can focus on the dynamics of the excited levels by projecting out the ground state
\begin{equation}
    \bar \rho_\mathrm{eff} = \check P \bar \rho,\quad \check P = \hat P\otimes\hat P^*,\quad \hat P = \sum_{l\neq 1} \ket{l}\bra{l}.
\end{equation}
Physically, this can be achieved by the postselection techniques~\cite{naghiloo2019, chen_quantum_2021}. The resulting dynamics of $\bar \rho_\mathrm{eff}$ is described by an effective Liouvillian superoperator
\begin{equation}
    \frac{d}{dt}\bar \rho_\mathrm{eff} = \BL_\mathrm{eff}\bar\rho,
\end{equation}
\begin{align}
    \BL_\mathrm{eff} &=\check P\BL\check P= \qty[(-i\hat H_\mathrm{eff})\otimes\mathbb1+\mathbb1\otimes(i\hat H_\mathrm{eff}^*)]\nonumber\\ &\hspace{2.4cm}+ \sum_{2\leqslant l<k}\Gamma_{lk}D\qty[\ket{l}\bra{k}],\\
    \hat H_\mathrm{eff} &= \sum_{l=2}^3\qty(\varepsilon_l-\frac{i\gamma_l}{2})\ket{l}\bra{l}\nonumber\\ &\hspace{1.5cm}+ \sum_{2\leqslant l< k} t_{lk}(\ket{l}\bra{k}+\ket{k}\bra{l}).
\end{align}
Here, we have denoted the decay rates into the ground state as $\gamma_l = \Gamma_{1l}$. 

Let us assume for simplicity, $\varepsilon_{2,3,4} = \varepsilon$, $t_{23}=t_{34}=t$, $t_{24}$ = 0. In the basis $\ket{2}$, $\ket{3}$, $\ket{4}$, the effective Hamiltonian takes the form
\begin{equation}
    \hat H_{\rm eff} = \begin{pmatrix}
        \varepsilon - \frac{i\gamma_2}{2} & t & 0\\
        t & \varepsilon - \frac{i\gamma_3}{2} & t\\
        0 & t & \varepsilon- \frac{i\gamma_4}{2}.
    \end{pmatrix}
\end{equation}
If $2\gamma_3 = \gamma_2+\gamma_4$ and $t^2 = (\gamma_4-\gamma_3)^2/2$,
it has a triple-degenerate eigenvalue $\varepsilon_\mathrm{EP} = \varepsilon - i\gamma_3/2$, which corresponds to the third-order EP3~\cite{shiralieva2025}. The minimal polynomial should look like $(\lambda - \varepsilon_\mathrm{EP})^n$ in this case, where $n$ is the size of the largest Jordan block. It is straightforward to check that $(\hat H_\mathrm{eff} - \varepsilon_\mathrm{EP})^n$ is nonzero for $n=2$ and gives a zero matrix for $n=3$, which confirms that it is really an EP3.

If we switch off the dissipation between the excited levels $\Gamma_{lk}\equiv0$, then the effective Liouvillian is determined by its Liouvillian part:
\begin{equation}
    \BL_\mathrm{eff} = \BL_\mathrm{eff}^\prime = \qty[(-i\hat H_\mathrm{eff})\otimes\mathbb1+\mathbb1\otimes(i\hat H_\mathrm{eff}^*)].
\end{equation}
Using the results of Refs.~\cite{shiralieva2025, trampus1966}, we can construct explicitly the Jordan basis of $\BL^\prime_\mathrm{eff}$ and determine its Jordan normal form as
\begin{equation}
    \BL_\mathrm{eff}^\prime \sim J_5(\lambda_{\mathrm{EP}})\oplus J_3(\lambda_\mathrm{EP})\oplus J_1(\lambda_\mathrm{EP}),
\end{equation}
where $\lambda_\mathrm{EP} = -\gamma_3$.

\begin{figure}
    \centering
    \includegraphics[width=0.5\linewidth]{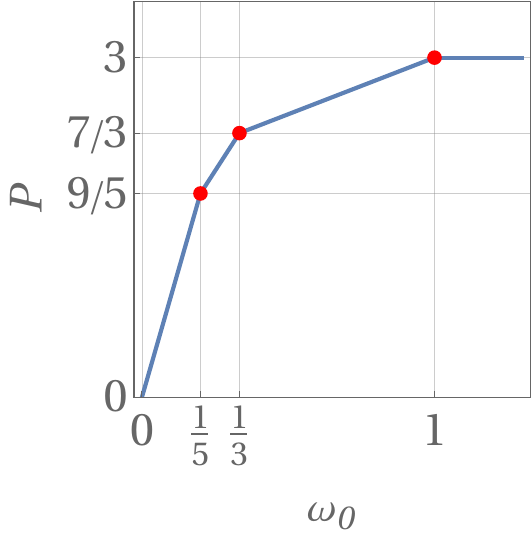}
    \caption{Tropicalized characteristic polynomial $\BP(\Gamma,\omega_0)$ of the effective Liouvillian $\BL_\mathrm{eff}$.}
    \label{fig:charpoly-liouvillian}
\end{figure}

As soon as we add the dissipation, the levels get split. To model a generic perturbation, we can pick the dissipation rates at random. As a concrete example, we set $\Gamma_{34} = \Gamma/13$, $\Gamma_{23} = 5\Gamma/8$ and $\Gamma_{24}= 39\Gamma/50$. The characteristic polynomial can be calculated using any symbolic computation system~\footnote{The coefficients of $\Gamma_{ij}$ were chosen rational, so that the characteristic polynomial can be computed exactly. Otherwise, one has to be careful about some of the polynomial coefficients getting evaluated to a very small number instead of exactly zero.}.  The resulting tropicalization of the characteristic polynomial is
\begin{multline}
    \BP(\Gamma, \omega_0) = \min\{9\omega_0, 8\omega_0+1, 7\omega_0+1,6\omega_0+1,\\5\omega_0+1,4\omega_0+1,3\omega_0+2,2\omega_0+2,\omega_0+2,3\},\\
    = \min{\{9\omega_0, 4\omega_0+1,\omega_0+2,3\}},
\end{multline}
which is plotted in Fig.~\ref{fig:charpoly-liouvillian}. As we see, there are roots of orders $1/5$, $1/3$ and $1$ with multiplicities $5$, $3$ and $1$, respectively.

\section{Conclusion}\label{sec:conclusion}

In this work, we have presented a systematic characterization of how the eigenvalues of non-Hermitian degeneracies split under generic and non-generic perturbations. Our analysis covers all possible degeneracy types, from $n$-bolical points to derogatory (multi-block) EPs, and thereby goes beyond the main focus of the literature on EPs of geometric multiplicity one. Employing techniques from tropical geometry, we have determined the asymptotic dispersion of NH degeneracies of any geometric multiplicity, together with the number of eigenvalues following each exponent, directly from the characteristic polynomial and without any matrix diagonalization. Applying this approach to $2 \times 2$, $3 \times 3$, $4 \times 4$ matrices, we have tabulated the complete splitting behavior of all Jordan structures, including anomalous exponents along non-generic directions that are not captured by generic-perturbation theory. We have further demonstrated the applicability of the approach across experimentally relevant settings, including EP-based sensing devices, as well as nonreciprocal systems, NH lattice models, and Liouvillian superoperators of open quantum systems.

A formulation that handles both generic and non-generic perturbations is a genuine asset for studying NH degeneracies. Under generic perturbations, the response is fixed entirely by the type of degeneracy being perturbed; under non-generic perturbations, the specific form of the perturbation also plays a critical role. This opens a wide range of possibilities for manipulating the degenerate structure of NH systems. In particular, a non-generic perturbation may partially lift the degeneracy, or it may leave the order of the degeneracy intact while changing its type~\footnote{For example, a type-(2,1)(2,1)
(2,1) EP can become a type-(3)(3)
(3) EP, the usual non-derogatory third-order EP.}. Converting one EP type into another offers a route to engineering the properties of NH systems. In our recent paper~\cite{Starkov2026}, we address precisely this question, determining which EP types can be converted into one another by infinitesimal perturbations through a hierarchy of degeneracy manifolds. Together, the two works cover the complete perturbative behavior of NH degeneracies: the present paper determines the splitting exponents when a degeneracy is lifted, while Ref.~\cite{Starkov2026} determines the accessible degeneracy types when it is not.

\section{Acknowledgment}
S.~S. acknowledges fruitful discussions and insightful communications with D.~Bahns, M.~Brandt, F.~Mohammadi and B.~Strumfels.
G.~S. is greatly indebted to I. Danilenko for the enlightening discussions, and would like to thank A. Shiralieva.
G.~S. is supported by DFG-SFB 1170 (Project-ID: 258499086) and EXC2147 ctd.qmat (Project-ID: 390858490).

 \bibliography{bibfile}

\end{document}